\documentclass[aps,amsmath,amssymb,floatfix,lengthcheck,preprintnumbers,prd, 
showpacs,superscriptaddress,twocolumn]{revtex4} 

\usepackage{hyperref}
\usepackage{verbatim}
\usepackage{graphicx}
\usepackage{color}
\usepackage{bm} % bold math
\usepackage{multirow}
\usepackage{ifpdf} % This lets us get figures right both in pdf and ps versions

\newcommand{\mb}{\overline{m}_b}
\newcommand{\mbmb}{ \mb(\mb)  }
\newcommand{\mbmbnf}{ \mb(\mb,n_f=3)  }

\newcommand{\msms}{\mbmb}
\newcommand{\MSbar}{\overline{MS  }}

\newcommand{\alphams}{\alpha_{\MSbar} }
\newcommand{\alphav}{\alpha_{V} }

\newcommand{\mpole}{M_b^{\rm pole  }}

\newcommand{\vegas}{\textsc{vegas} }
\newcommand{\hippy}{\textsc{HiPPy} }

\newcommand{\hpsrc}{\textsc{HPsrc} }
\newcommand{\taylur}{\textsc{TaylUR} }

\newcommand{\fortran}{\textsc{FORTRAN} }
\newcommand{\MS}{{\overline{MS}}}
\renewcommand{\vec}[1]{{\bm{#1}}}
\newcommand{\ONE}{\makebox[0pt][l]{\hspace{0.05cm}1}1}

\newcommand{\half}{\frac{1}{2}}
\newcommand{\Tr}{\mbox{Tr}}

\newcommand{\bs}{\mbox{\boldmath $s$}}
\newcommand{\bx}{\mbox{\boldmath $x$}}
\newcommand{\by}{\mbox{\boldmath $y$}}

\newcommand{\bk}{\mbox{\boldmath $k$}}
\newcommand{\bp}{\mbox{\boldmath $p$}}
\newcommand{\bfp}{{\bf p}}
\newcommand{\bq}{\mbox{\boldmath $q$}}
\newcommand{\bn}{\mbox{\boldmath $n$}}
\newcommand{\bsn}{{\mbox{\boldmath $\scriptstyle n$}}}
\newcommand{\bsp}{{\mbox{\boldmath $\scriptstyle p$}}}

\newcommand{\bsx}{{\mbox{\boldmath $\scriptstyle x$}}}
\newcommand{\bsq}{{\mbox{\boldmath $\scriptstyle q$}}}
\newcommand{\bsy}{{\mbox{\boldmath $\scriptstyle y$}}}
\newcommand{\be}{\mbox{\boldmath $e$}}
\newcommand{\tU}{\widetilde{U}}
\newcommand{\cP}{{\cal P}}
\newcommand{\cL}{{\cal L}}
\newcommand{\la}{\langle}
\newcommand{\ra}{\rangle}
\newcommand{\tG}{\widetilde{G}}
\newcommand{\hG}{\hat{G}}
\newcommand{\tK}{\widetilde{K}}
\newcommand{\tGamma}{\widetilde{\Gamma}}

\newcommand{\hq}{\hat{q}}

\newcommand{\hqzero}{\hat{q_0}}
\newcommand{\ben}{\begin{equation}}
\newcommand{\een}{\end{equation}}
\newcommand{\bea}{\begin{eqnarray}}
\newcommand{\eea}{\end{eqnarray}}
\newcommand{\ba}{\begin{array}}
\newcommand{\ea}{\end{array}}
\newcommand{\Di}{\displaystyle}
\newcommand{\BES}{\begin{eqnarray*}}
\newcommand{\EES}{\end{eqnarray*}}
\newcommand{\nn}{\nonumber}
\newcommand{\cambridge}{Department~of~Applied~Mathematics~and~Theoretical~Physics, 
University~of~Cambridge, Centre~for~Mathematical~Sciences, Cambridge~CB3~0WA, United~Kingdom}
\newcommand{\glasgow}{SUPA, School~of~Physics~and~Astronomy, Kelvin
Building, University~of~Glasgow, 
Glasgow~G12~8QQ, Scotland}

\begin{document}

\title{The mass of the $b$-quark from lattice NRQCD and lattice perturbation theory}

\author{A.J. \surname{Lee}}
\affiliation{\cambridge}
\author{C.J. \surname{Monahan}}
\affiliation{\cambridge}
\affiliation{Physics Department, College of William and Mary, Williamsburg, Virginia 23187, USA}
\author{R.R. \surname{Horgan}}
\affiliation{\cambridge}
\author{C.T.H. \surname{Davies}}
\affiliation{\glasgow}
\author{R.J. \surname{Dowdall}}
\affiliation{\cambridge}
\affiliation{\glasgow}
\author{J. \surname{Koponen}}
\affiliation{\glasgow}

%\preprint{DAMTP-2011-....}
%\preprint{GLA-....}
\collaboration{HPQCD collaboration}
\homepage{http://www.physics.gla.ac.uk/HPQCD}

\begin{abstract}

We present a determination of the $b$-quark mass accurate through $\mathcal{O}(\alpha_s^2)$ in perturbation theory and including partial contributions at $\mathcal{O}(\alpha_s^3)$. Nonperturbative input comes from the calculation of the $\Upsilon$ 
and $B_s$ energies in lattice QCD including the effect of $u$, $d$ and $s$ sea quarks. We use an improved NRQCD action for the $b$-quark. This is combined with the heavy quark energy shift in NRQCD determined using a mixed approach of high-$\beta$ simulation and automated lattice perturbation theory. Comparison with experiment enables the quark mass to be extracted: in the $\MS$ scheme we find $\mbmb$ = 4.166(43) GeV.

\end{abstract}

\maketitle

%%%%%%%%%%%%%%%%%%%%%%%%%%%%%%%%%%%%%%%%%%%%%%%%%%%%%%%%%%%%%%%%%%%%%%%%%%%
%%%%%%%%%%%%%%%%%%%%%%%%%%%%%%%%%%%%%%%%%%%%%%%%%%%%%%%%%%%%%%%%%%%%%%%%%%%
\section{\label{sec:intro}Introduction}
%%%%%%%%%%%%%%%%%%%%%%%%%%%%%%%%%%%%%%%%%%%%%%%%%%%%%%%%%%%%%%%%%%%%%%%%%%%
%%%%%%%%%%%%%%%%%%%%%%%%%%%%%%%%%%%%%%%%%%%%%%%%%%%%%%%%%%%%%%%%%%%%%%%%%%%

The accurate determination of quark masses is an important component of 
high-precision tests of the Standard Model. Because quarks cannot be isolated 
experimentally, the mass must be defined carefully and its extraction from quantities 
that are accessible to experiment must be well controlled from the theory side. The 
$b$-quark mass is particularly important:  its uncertainty feeds into errors in tests 
of the Standard Model in B physics as well as into the cross-section for the Higgs 
decay, $H \rightarrow b\overline{b}$.

The most accurate results to date for the $b$-quark mass come from comparison of the 
experimental cross section for $e^+e^-$ to hadrons in the bottomonium region with 
high-order ($\alpha_s^3$) continuum QCD perturbation theory 
\cite{Chetyrkin:2009fv,Narison:2011rn,Narison:2011xe}. Errors of 0.5\% are possible. A 
similar method has now been applied to lattice QCD results 
\cite{Allison:2008xk,mcneile10}, using pseudoscalar correlators made from heavy 
quarks instead of the experimental cross-section. For these calculations, the 
experimental input is the value of the meson mass (in this case the $\eta_b$) used to 
tune the lattice $b$-quark mass. Again a 0.5\% error is achieved and good agreement 
is seen with the continuum results.

It is important to test these determinations against a different method of obtaining 
the $b$-quark mass which has completely uncorrelated systematic errors. This is the 
aim of this paper. We use a direct determination from full lattice QCD calculations 
of the binding energy of both $\Upsilon$ and $B_s$ mesons. Since we use a 
nonrelativistic effective theory for the $b$-quark (NRQCD) \cite{Thacker:1990bm,Lepage:1992} this needs a calculation 
of the heavy quark energy shift. We do this in lattice QCD perturbation theory 
through two-loops (with partial three-loop contributions), significantly improving on 
earlier determinations that used one-loop calculations \cite{gray05}. We have also 
implemented a one-loop improved NRQCD action to reduce systematic errors.

Calculating higher order loop corrections in lattice perturbation theory for heavy 
quarks in NRQCD grows ever more difficult with each order owing to the increasing 
number of diagrams and the complicated vertex structure. Various authors 
\cite{dimm95,hart04,mueller09b} have suggested an approach in which the heavy quark 
propagator is measured in the weak coupling regime and the renormalization parameters 
are fitted to a polynomial in $\alpha_s$, thus obtaining the radiative corrections 
beyond one loop. This method is certainly practical for obtaining the quenched 
contributions to renormalization parameters since quenched gauge configurations are 
relatively cheap to generate. At two loop order there are relatively few remaining 
diagrams with sea quark loops and these can be feasibly computed using automated 
lattice perturbation theory. In contrast, there are many two loop diagrams containing 
only gluon propagators that pose a challenging task for direct evaluation with 
automated lattice perturbation theory. We therefore employ a mixed approach to the 
determination of the two-loop heavy quark energy shift combining quenched 
high-$\beta$ calculations with automated lattice perturbation theory for the sea 
quark pieces.

In Section \ref{sec:mb-extract} we discuss how we extract the $b$-quark mass from 
simulations of lattice NRQCD. Section \ref{sec:auto-lpt}
describes the automated lattice perturbation theory computation of the fermionic 
contributions to the two loop energy shift. We present our implementation of the 
high-$\beta$ method in Section \ref{sec:hbeta-method} including the concomitant 
finite volume perturbation theory in Appendix \ref{sec:finite_v_pth}. The details of 
the standard non-perturbative part of the calculation are given in Section 
\ref{sec:nonpert}. Finally we detail the extraction of the $\MS$ mass in Section 
\ref{sec:mb-msbar} and present our conclusions in Section \ref{sec:conclusion}.

\section{\label{sec:mb-extract} Extracting the \texorpdfstring{$b$}{b}-quark mass}
Quark confinement ensures that quark masses are not physically measurable
quantities, so the notion of quark mass is a theoretical construction. A
wide range of quark mass definitions exist, often tailored to exploit the
physics of a particular process. One common choice of quark mass is the
pole mass, defined as the pole in the renormalised heavy quark propagator.
The pole mass, however, is a purely perturbative concept and suffers from
infrared ambiguities known as renormalons
\cite{bigi94b,beneke94a}. 
A better mass is the running mass in the $\MS$ scheme, which is free of renormalon
ambiguities by construction, and is the usual choice for quoting the quark masses. 
Lattice calculations use the renormalon-free bare lattice mass which must then be matched to $\MS$ to enable meaningful comparison. 
We match bare lattice
quantities to the $\MS$ mass using the pole mass as an intermediate step.
Any renormalon ambiguities cancel in the full matching procedure between
the lattice quantities and the $\MS$ mass, as we argue below.
For an explicit demonstration, see \cite{Bodwin:1998mn}.

\subsection{Extracting the pole mass}\label{ssec:pmass}

We determine the heavy quark pole mass,
$M_{pole}$, by relating it to the experimental $\Upsilon$ mass
$M_{\Upsilon}^{\mathrm{expt}}$.
The mass of a heavy meson is given by twice the pole quark mass plus the binding energy.
In an effective theory such as NRQCD, physics above the scale of the $b$-quark mass is removed and the zero of energy for the heavy quark is shifted by $E_0$, leading to the relation \cite{davies94}: 
\begin{equation}\label{eq:e0}
2M_{pole} = M_{\Upsilon}^{\mathrm{expt}} - a^{-1}(aE_{\mathrm{sim}} -
2aE_0).
\end{equation}
Here $E_{\mathrm{sim}}$ is the energy of the $\Upsilon$ meson at zero
momentum, extracted from lattice NRQCD data at lattice spacing $a$. The
quantity
$(E_{\mathrm{sim}} - 2E_0)$ corresponds to the ``binding energy'' of
the
meson in NRQCD and we must determine $E_0$ perturbatively  in order to find $M_{pole}$.
With our NRQCD action, we can also calculate the pole mass using the $B_s$ meson
\begin{equation}\label{eq:e0Bs}
M_{pole} = M_{B_s}^{\mathrm{expt}} - a^{-1}(aE_{\mathrm{sim}}^{B_s} -
aE_0).
\end{equation}
We use this as a check for systematic errors which could be quite different in heavy-heavy and heavy-light systems.

In principle one could extract the quark mass by directly matching the pole
mass to the bare lattice NRQCD mass in physical units, $m_0$, via the heavy
quark mass renormalisation, $Z_{m_0}$, 
\begin{equation}\label{eq:zmlatt}
M_{pole} = Z_{m_0}(am_0) m_0.
\end{equation}
We found, however, that extracting a sufficiently precise quenched two loop
mass renormalisation from high-$\beta$ simulations was not possible with
the statistics available. In this paper, we therefore discuss only the
energy shift method.

\subsection{\label{ssec:pole_msbar}\texorpdfstring{Matching the pole mass to the $\MS$
mass}{Matching the pole mass to the MS bar mass}}

The mass renormalisation relating the pole mass to the $\MS$ mass,
$\mb$, evaluated at some scale $\mu$, is given by
\begin{equation}\label{eq:zmcont}
\mb(\mu) = Z_{M}^{-1} (\mu)M_{pole},
\end{equation}
and has been calculated to three-loops in \cite{Melnikov:2000qh}.

Although the pole mass is plagued by renormalon ambiguities, these
ambiguities cancel when lattice quantities are related to the $\MS$
mass. This can be seen by equating Eqns
(\ref{eq:e0}) and (\ref{eq:zmlatt}) and rearranging them to obtain
\begin{equation}
2(Z_{m_0} m_0 - E_0) = M_{\Upsilon}^{\mathrm{expt}} -E_{\mathrm{sim}}.
\end{equation}
The two quantities on the right hand side of the equation are renormalon
ambiguity free: $M_{\Upsilon}^{\mathrm{expt}}$ is a physical quantity 
and $E_{\mathrm{sim}}$ is determined nonperturbatively from lattice
simulations. Any renormalon ambiguities in the two power series, $Z_{m_0}$ and
$E_0$, on the left-hand side of the equation must therefore cancel at every order in $\alpha_s$. 
This renormalon cancellation is also evident in the direct matching of the
bare lattice mass to the $\MS$ mass,
\begin{equation}
\mb(\mu) = Z_{m_0}(am_0) Z_{M}^{-1} (\mu)m_0,
\end{equation}
as both $\mb$ and $m_0$ are renormalon-free.

We combine Eqns \eqref{eq:e0} and \eqref{eq:zmcont} to relate lattice
quantities to the $\MS$ mass
\begin{align}
\label{eq:msbar}
\mb(\mu) =    \frac{1}{2}Z_{M}^{-1} (\mu) 
 \left[M_{\Upsilon}^{\mathrm{expt}} -
a^{-1}(aE_{\mathrm{sim}} - 2aE_0)\right],
\end{align}
and similarly for the $B_s$ meson
\begin{align}
\label{eq:msbar_Bs}
\mb(\mu) = {} &  Z_{M}^{-1} (\mu) 
   \left[M_{B_s}^{\mathrm{expt}} -
a^{-1}(aE_{\mathrm{sim}, B_s} - aE_0)\right].
\end{align}
These relations will be used to extract $\msms$ once we have calculated $E_0$ and $E_{\rm sim}$, which we describe in detail in the next sections.

\subsection{\label{ssec:action}NRQCD, gluon and light quark actions}

We now describe the heavy quark, gluon and light quark actions used in our calculation.
We use the Symanzik improved $\mathcal{O}(v^4)$ NRQCD action, given in \cite{gray05,dowdall12}, which has already been successfully used by HPQCD in a number of heavy quark physics calculations, see e.g. \cite{gray05,dowdall12,Daldrop:2011aa,Dowdall:2012ab,Gregory:2010gm,fBpaper}. 
The Hamiltonian is given by
\begin{eqnarray}
 aH   &=& aH_0  + a\delta H;  \\
 aH_0 &=& -  \frac{\Delta^{(2)}}{2 am_0},  \\
a\delta H
&=& 
- c_1 \frac{(\Delta^{(2)})^2}{8( am_0)^3}
+ c_2 \frac{ig}{8(am_0)^2}\left(\bf{\nabla}\cdot\tilde{\bf{E}}\right. -
      \left.\tilde{\bf{E}}\cdot\bf{\nabla}\right) \nonumber \\
& & 
- c_3 \frac{g}{8(am_0)^2} \bf{\sigma}\cdot\left(\tilde{\bf{\nabla}}\times\tilde{\bf{E}}\right. -
      \left.\tilde{\bf{E}}\times\tilde{\bf{\nabla}}\right) \nonumber \\
& & 
- c_4 \frac{g}{2 am_0}\,{\bf{\sigma}}\cdot\tilde{\bf{B}}  
+ c_5 \frac{a^2\Delta^{(4)}}{24 am_0} \nonumber \\
& & 
-  c_6 \frac{a(\Delta^{(2)})^2}{16n(am_0)^2} .
\label{deltaH}
\end{eqnarray}
$\Delta^{(2)}, \nabla$ and $\Delta^{(4)}$ are covariant lattice derivatives, $\tilde{\bf{E}}$ and $\tilde{\bf{B}}$ are improved chromo-electric and magnetic field strengths, $n$ is a stability parameter that will be described below and $am_0$ is the bare $b$-quark mass in lattice units. The $c_i$ are the Wilson coefficients of the effective theory and the terms are normalised such that they have the expansion $c_i = 1 + \alpha_s c_i^{(1)} + \mathcal{O}(\alpha_s^2)$. All gauge fields are tadpole improved with the fourth root of the plaquette $u_{0,P}$.

The one loop corrections $c_i^{(1)}$ are described in \cite{dowdall12} and we include these for $c_1,c_4,c_5,c_6$ in the high-$\beta$ simulation and the nonperturbative determination of $E_{\rm sim}$. The $c_i^{(1)}$ are a function of the effective theory cutoff, in this case the bare quark mass $am_0$, but the total coefficient will also depend on the scale for $\alpha_s$. We estimate the appropriate scale for several of the coefficients using the BLM procdure \cite{Brodsky:1982gc} which gives $q^*=1.8/a$ for $c_1,c_6$ and $q^*=1.4/a$ for $c_5$. For $c_4$ we take $q^*=\pi/a$.
The values of the one loop corrections for two bare masses relevant to this calculation are given in table \ref{tab:ci}. We use $\alpha_s$ in the $V$-scheme.

\begin{table}
\caption{ Values of the one loop corrections in the series' $c_i = 1.0 + \alpha_s c_i^{(1)}$ at two bare masses, and the scale at which each coefficient is evaluated.
}
\label{tab:ci}
\begin{ruledtabular}
\begin{tabular}{llll}
 & $c^{(1)}_i$  & $c^{(1)}_i$ &  \\
Coefficient & $am_0=2.5$   & $am_0=1.72$ & $q^*$  \\
\hline
$c_1$ & 0.95 & 0.766  & $1.8/a$ \\
%$c_2$ & 0.0  & 0.0    &  - \\
%$c_3$ & 0.0  & 0.0    &  - \\
$c_4$ & 0.78 & 0.691  & $\pi/a$ \\
$c_5$ & 0.41 & 0.392  & $1.4/a$ \\
$c_6$ & 0.95 & 0.766  & $1.8/a$ \\
\end{tabular}
\end{ruledtabular}
\end{table}

The $b$-quark propagators are generated by time evolution using the equation
\begin{multline}
G(\vec{x},t+1) = 
	\left( 1-\frac{a\delta H}{2}\right)\left(1-\frac{aH_0}{2n}\right)^nU^{\dag}_{t}(x) 
	 \\
         \times \left(1-\frac{aH_0}{2n}\right)^n\left(1-\frac{a\delta H}{2}\right) G(\vec{x},t) 
\label{eq:evol}
\end{multline}
for some initial condition $G(\bx,0)$. The parameter $n$ is included for numerical stability and is set to 4, which is sufficient for all quark masses used here. Once it is high enough, results do not depend on the value of $n$ \cite{gray05}. 

The gluon action is a Symanzik improved L\"uscher-Weisz action \cite{luscher85,alford95}
\bea
S_{LW} [ U ] & = &  \displaystyle \beta_{ pl } \sum_{ x }\frac{ 1 }{ N_c }  \,
\mbox{Re Tr} \, ( \ONE - U_{ pl } ) \nn \\
& + &  \displaystyle \beta_{ rt } \sum_{ x }\frac{ 1 }{ N_c }  \, \mbox{Re Tr}
\, ( \ONE - U_{ rt } ) \nn \\
& + &  \displaystyle \beta_{ pg } \sum_{ x  }\frac{ 1 }{ N_c }  \, \mbox{Re Tr}
\, ( \ONE - U_{ pg } ),
\eea
where
\bea
\beta_{ pl } & = &\displaystyle \frac{ 10 }{ g^2 }, \\
\beta_{ rt } & = & \displaystyle -\frac{ \beta_{ pl } }{ 20
u_{0,P}^{  2 } } \, \left( 1 + 0.4805  \alpha_s \right), \\
 \beta_{ pg } & =  & \displaystyle -\frac{ \beta_{ pl }
}{ u_{0,P}^{ 2 } } \, 0.03325 \alpha_s.
\eea
$u_{0,P}$ is the tadpole improvement factor coming from the fourth-root of the plaquette. The same action is used for the MILC gauge configurations used in the non-perturbative determination of $E_{\rm sim}$ and for the high-$\beta$ simulations. The action in the high-$\beta$ simulations includes an additional factor coming from the use of twisted boundary conditions, see section \ref{sec:hbeta-method}. 
The value of $\alpha_s$ used in the 
improvement coefficients is given by the formula used by the MILC collaboration \cite{Bazavov:2009bb}:
\begin{equation}
\alpha_s = 1.3036\log(u_{0,P}(\beta)).
\end{equation}
Here we use the quenched values of $u_{0,P}(\beta)$ determined from our high-$\beta$ configurations. The MILC configurations used in 
our nonperturbative analysis include sea quarks and so have 
additional $O(n_f\alpha_s^2)$ contributions. However, these only 
affect $E_0$ at  $O(n_f\alpha_s^3)$ and so appear in terms we have 
not calculated anyway. These terms are part of our error budget.
We give more details of the
generation of high-$\beta$ configurations in
Appendix \ref{sec:generating_configurations}.

Light sea quarks are included with the ASQtad improved staggered action \cite{Lepage:1998vj} in both the $n_f=2+1$ MILC gauge configurations used to determine $E_{\rm sim}$ \cite{Aubin:2004wf,Bazavov:2009bb} and in the automated perturbation theory for $E_0$. 

\section{\label{sec:pert_theory}Perturbative determination of the heavy quark energy shift }
Here we first describe the calculation of the one-loop contribution and the two-loop fermionic contribution
to $E_0$. 
The high-$\beta$ method used to compute the gluonic two-loop contribution is described in the section \ref{sec:hbeta-method}.

\subsection{\label{sec:auto-lpt}Automated lattice perturbation theory}
We calculate the one loop gluonic and the two loop sea quark contributions to the heavy quark 
renormalization constants using the automated lattice perturbation theory routines 
\hippy and \hpsrc \cite{Hart:2004bd,Hart:2009nr}. These routines have now been widely 
used and extensively tested in a variety of perturbative calculations, for example in 
\cite{drummond02,drummond03a,drummond03b,hart04,hart07,mueller11, 
hammant11,dowdall12}.

Evaluating the relevant Feynman integrals with \hippy and \hpsrc is a two-stage 
process: firstly the \verb+python+ routine \hippy generates Feynman rules encoded in 
``vertex files''. These vertex files are then read in by the \hpsrc code, a 
collection of \fortran modules that reconstruct the diagrams and evaluate the 
corresponding integrals numerically, using the \vegas algorithm \cite{Lepage:1977sw}. 
All derivatives of the self energy are implemented analytically using the derived 
\verb+taylor+ type, defined as part of the \taylur package \cite{hippel10}. Our 
computations of these diagrams were performed on the Darwin cluster at the Cambridge 
High Performance Computing Service with routines adapted for parallel computers using 
Message Passing Interface (MPI).

There are several advantages associated with using automated lattice perturbation 
theory, and the \hippy/\hpsrc routines in particular. First, automation removes the 
need to manipulate complicated expressions by hand. Secondly, the modular nature of the \hippy 
and \hpsrc routines greatly simplifies the use of different actions. Once Feynman 
diagrams are encoded in an \hpsrc routine, the same calculation can be easily 
repeated with different quark and gluon actions by simply changing the input vertex 
files. This allows one to relatively easily reproduce previously published results 
for different actions, which serves as a nontrivial check of the routines.

Furthermore, the modules in \hpsrc can be reused. We took advantage of this for the 
two loop calculations presented in this paper: the same fermionic insertions in the 
gluon propagator appear in the two loop diagrams for both the heavy quark energy 
shift and the tadpole improvement factor, $u_0$. 

\begin{figure}
\begin{centering}
\includegraphics[width=0.7\hsize]{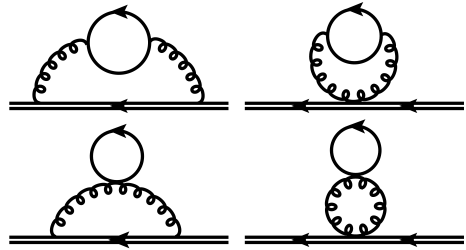}
\caption{\label{fig:fermionE0}Fermionic
contributions to $E_0$, calculated using automated lattice perturbation theory. 
Double lines indicate heavy quarks, curly lines are gluons and single lines represent light sea quarks.}
\end{centering}
\end{figure}

We wrote two ``skeleton'' one loop \hpsrc routines: one to calculate the one loop 
energy shift and one for the one loop tadpole improvement factor. Reproducing 
previously published results, such as those in \cite{gulez04} and \cite{Nobes:2001} 
respectively, confirmed that these one loop routines were correct. The corresponding 
two loop diagrams (see Figure \ref{fig:fermionE0}) are simply the one 
loop skeleton diagrams with the ``bare'' gluon propagator replaced by the ``dressed'' 
gluon propagator that includes the fermion insertions; these insertions were 
calculated in a separate routine \verb+gluon_sigma+. This routine was debugged by confirming that the 
appropriate Ward identity was satisfied by the dressed gluon propagator. 

At two loops there are four diagrams with internal fermions that
contribute to the energy shift. We illustrate these
contributions in Figure \ref{fig:fermionE0}.
Double lines are heavy quark propagators coming from the improved NRQCD action, single lines are ASQtad sea quark propagators and curly lines are from the Symanzik improved gluon action.
The radiative corrections to the NRQCD and ASQtad actions described in section \ref{ssec:action} are not included in the perturbative calculation as these only affect $E_0$ at higher order in $\alpha_s$.

We calculated the heavy quark energy shift at two different heavy quark masses
discussed in section \ref{sec:nonpert}.
At each heavy quark mass we use 
nine different light quark masses and extrapolate to zero light quark mass. We 
tabulate our extrapolated results in Table \ref{tab:E0_pertres} where they appear as the 
$n_f$-dependent contribution to $E_0^{(2)}$.

The energy shift is infrared finite, but we introduce a gluon mass as an intermediate 
regulator to ensure convergence for the numerical integration. We 
confirmed that the results are independent of the gluon mass for sufficiently small 
gluon mass, which in this case was approximately $a^2\lambda^2 < 10^{-6}$.

We will also need the sea quark contribution to the tadpole improvement factor $u_0$ since the high-$\beta$
simulation includes only the gluonic piece. We calculate this using the automated perturbation theory.
The perturbative expansion for the tadpole factor is written as
\ben
u_0 = 1 - u_0^{(1)}\alpha_L - u_0^{(2)}\alpha_L^2 + {\cal O}(\alpha_L^3).
\een
The two loop expansion for the plaquette tadpole is given by Mason \cite{Mason:2004aa}
and we explicitly computed the one-loop coefficient and the two-loop $n_f$ coefficient 
which we quote here and which both agree with Mason. The result is
\begin{multline}
u_{0,P} = 1 - 0.76708(2)\alpha_L  
\\ - (1.7723-0.069715(7)n_f)\alpha_L^2 + {\cal O}(\alpha_L^3).
\end{multline}
We require only the coefficent of $n_f\alpha_L^2$. For completeness we also computed 
the two-loop $n_f$ contribution to the Landau tadpole. The quenched two-loop Landau tadpole
was computed by Nobes et al. \cite{Nobes:2001} and together with our result the Landau tadpole is
\begin{multline}
u_{0,L} = 1 - 0.7501(1)\alpha_L \\ - (2.06(1)-0.0727(1)n_f)\alpha_L^2 + {\cal O}(\alpha_L^3).  
\end{multline}

\subsection{\label{sec:hbeta-method}The high\texorpdfstring{-$\beta$}{\ beta} method}

The high-$\beta$ method allows us to compute the gluonic contributions to the quark 
propagator by generating an ensemble of quenched lattice gauge configurations at very weak 
coupling and calculating the dressed $b$-quark propagator. The energy of the propagator 
can then be described very well by a power series in the QCD coupling, which we fit to the 
Monte-Carlo data to extract the relevant two-loop and higher contributions to $E_0$.

It is important in high-$\beta$ studies to eliminate non-perturbative contributions which 
are due to the tunnelling of fields and their associated Polyakov lines, or torelons, 
between $Z_3$ vacua associated with toron gauge configurations \cite{Luscher:1982ma}. Such 
tunnelling is suppressed using twisted boundary conditions 
\cite{'tHooft:1979uj,Luscher:1986,Trottier:2001vj} for which there is no 
zero mode for the non-abelian gauge field. The Polyakov line that traverses all the 
directions with twisted boundary conditions has a non-zero expectation value for a given 
configuration. This expectation value is complex and if no tunnelling has occured it is 
proportional to an element of $Z_3$. We verify that this is the case for the 
configurations we use. As is shown later in this section, see the discussion leading to Eqs. 
(\ref{eq:Rmatrix}) (\ref{eq:Rmatrix1}), twisted boundary conditions also considerably reduce 
finite-size, $L$-dependent effects which significantly aids the fitting process.

We carry out the high-$\beta$ simulation on finite size lattices of volume $L^3 \times T$, 
with typically $T=3L$, for a range of values for $\beta$ and $L$. Here $L$ is the spatial 
extent and $T$ the temporal extent of the lattice. We use $L$ values from 3 to 10 
inclusive and $\beta_{pl}$ values of 12,15,16,20,24,27,32,38,46,54,62,70,80,92, and 120. We then 
perform a simultaneous fit in $\alpha_s$ and $L$ to deduce the $L \to \infty$ limit for 
the expansion of measured quantities as a power series in $\alpha_s$.

We denote the gauge fields by $U_\mu(x)$, and on a lattice with $L_\mu$ sites in the $\mu$ 
direction they satisfy the boundary condition
\ben
U_\mu(x+L_\nu\be_\nu) = \Omega_\nu U_\mu(x) \Omega^\dagger_\nu\;,
\een
where the twist matrices are defined by
\bea
\Omega_\mu \Omega_\nu&=&z^{n_{\mu\nu}}\Omega_\nu \Omega_\mu, \nn\\
z &=& \exp\left({2\pi i/N_c}\right),~~n_{\mu\nu} \in (0,\ldots,N_c-1)\;.
\label{eqn:omegas_and_z}
\eea
Here $n_{\mu\nu}$ is antisymmetric and its values must be chosen so that 
$\epsilon_{\mu\nu\sigma\rho}n_{\mu\nu}n_{\sigma\rho} = 0|_{N_c}$. This choice ensures 
configurations have zero topological charge. For $N_c=3$ we apply a non-trivial twist in 
the spatial directions, which we label 1, 2 and 3, with $n_{12} = n_{13} = n_{23} = 1$ and 
$n_{\mu 4} = 0$.

With twisted boundary conditions, the fermion fields, $\psi$, are $N_c \times N_c$ 
colour-times-smell matrices. ``Smell'' is a new quantum number that allows twisted 
boundary conditions to be applied to fermion fields; colour labels the rows and smell the 
columns. Then, as for the gauge fields,
\ben
\psi(x+L_\nu \be_\nu) = \Omega_\nu \psi(x) \Omega^\dagger_\nu\;.
\een

Under a gauge transformation given by the $SU(N_c)$ field $g(x)$ the quantum fields
transform as
\bea
U_\mu(x) &\to& g(x)U_\mu(x)g^\dagger(x+\be_\mu), \nn\\
\psi(x) &\to& g(x)\psi(x), \label{eq:gt}
\eea
where $g(x+L_\nu\be_\nu) = \Omega_\nu g(x) \Omega^\dagger_\nu$. We define the auxiliary gauge fields 
\ben
\tU_\mu(x) = \left\{
\ba{lcl}\Di U_\mu(x)&~~&\Di x_\mu \ne L_\mu, \\
          \Di U_\mu(x)\Omega_\mu&~~&\Di x_\mu = L_\mu\;. 
\ea \right. 
\een
Then under a gauge transformation $\tU_\mu(x)$ transforms as in Eq. (\ref{eq:gt}) but now
with $g(x)$ regarded as periodic: $g(x+L_\mu\be_\mu) = g(x)$.

The gauge action is of the form 
\ben
S(U) = \beta \sum_{P;x \in \Lambda} c_P f_P(x) P(\tU,x)\;,
\een
where $\Lambda$ is the set of all lattice sites; $P(\tU,x)$ is the trace
over a general Wilson loop;
$c_P$ is a numerical
coefficient and $f_P(x) \in Z_{N_c}$ is a phase factor defined by
\ben
f_P(x) = \prod_{\mu < \nu}
\left(z^{n_{\mu\nu}}\right)^{-\omega_{\mu\nu}(P,x)}\;.
\een
Here $\omega_{\mu\nu}(P,x)$ is the winding number of the Wilson loop projected onto the 
$(\mu,\nu)$ plane about the point $x_\mu = x_\nu = (L+1/2)$. An explicit representation 
for the twist matrices $\Omega_\mu$ is not needed to compute $f_P(x)$. When fermions are 
included, however, the implementation of twisted boundary conditions for general Wilson 
lines does require a representation for the $\Omega_\mu$ to be chosen.

One method for implementing the boundary conditions extends the lattice by tiling with 
twisted periodic translations of the original configuration, effectively surrounding the 
lattice with a halo of links. This method has major disadvantages: it is difficult to 
parallelize because the physical sites are a subset of the tiled lattice array; it 
requires more storage; and in improved NRQCD the Wilson lines can extend far into the 
tiled region, which means that the extent of the halo needs to be significant. Rather than 
extending the lattice we write the action in terms of the auxiliary gauge fields, 
$\tU_\mu(x)$. Then one can show that all Wilson lines can be constructed using the 
auxiliary gauge fields with periodic boundary conditions multiplied on the right by an 
$SU(N_c)$ matrix. This $SU(N_c)$ matrix, which we denote $R(\cP)$, is constructed from a 
product of the twist matrices, $\Omega_\mu$, and is determined by the ordered and signed 
sequence in which the line crosses the boundaries. We now discuss this construction in 
more detail.

A general path $\cP(x,y;\bs)$ starting at site $x$ on a lattice in dimension $D$ is
defined by an ordered list $\bs = [s_0,s_1, \ldots s_{l-1}]$  of signed integers,
$s_i,~1 \leq |s_i| \le D$, which denote the steps along the path. The $j$-th point on the path
is $z_j$ where
\ben
z_0 = x,~~z_{j+1} = z_j+\be_{s_j}~~0 \le j < l\;,
\een
with endpoint defined by $y = z_l$. We define the ordered product of links along the path
$\cP(x,y;\bs)$ to be
\ben
\cL(\tU;\cP) = \left[{\cal T} \prod_{i=0}^{l-1} \tU_{s_i}(z_i)\right],
\een
where, for $\mu \in \{1,2, \ldots, D\}$,
\ben
\tU_{-\mu}(x) = \tU^\dagger_\mu(x-\be_\mu),~~~\be_{-\mu} = -\be_\mu,
\een
and the $\tU$ fields satisfy the periodic boundary condition
\ben
\tU_\mu(x+L_\nu\be_\nu) = \tU_\mu(x)\;.
\een
The ordering operator $\cal T$ means that matrices in the product are ordered from left to right
with increasing index, $i$. The Wilson line $L(\tU;\cP)$ associated with the path $\cP(x,y;\bs)$ 
is then
\ben
L(\tU;\cP) = \cL(\tU;\cP)R(\cP)\;.
\een
To implement the twisted boundary conditions without using a lattice halo we define the
$SU(N_c)$ matrix as follows. A list $[c_0(x),c_1(x), \ldots, c_{p-1}(x)]$ is associated with 
the Wilson line starting at $x$, where the $c_j$ are signed integers $1 \le |c_j| \le D$.
The line crosses a boundary of the hypercube $p$ times. On the $j$-th crossing it crosses a
boundary in a direction parallel to the $\mu_j$ axis in the positive (negative) direction.
We define the corresponding $c_j$ to be $c_j = -\mu_j(\mu_j)$. $R(\cP)$ is then given by
\ben
R(\cP)~=~\left({\cal T}\prod_{j=0}^{p-1} \Omega_{c_j}\right)^\dagger\;,
\een
with the convention $\Omega_{-\mu} = \Omega^\dagger_\mu$ and where $\cal T$ is the index-ordering
operator defined above. $L(\tU;\cP)$ is then the parallel transporter from the endpoint $y$
back to the start point $x$.  By expressing the Wilson line in terms of the $\tU$ fields the
boundary conditions are implemented simply by right-multiplication by $R(\cP)$. A similar
result holds for the evolution of the NRQCD Green function as we will describe below. With these
conventions, a Wilson loop $W(x,\bs)$, located at $x$ and defined by the path $\cP(x,x;\bs)$, is
given by
\ben
W(x,\bs)~=~\frac{1}{N_c} \Tr \left( L(\tU;\cP(x,x;\bs)) \right)\;.
\een

The basis states for the fermion field $\psi$ are the $N_c^2$ independent
$N_c \times N_c$, colour-times-smell, real matrices. Twisted boundary conditions admit
fractional momenta on the lattice and for twisted boundary conditions in the $1,2,3$ directions
and periodic boundary conditions in the fourth direction, the allowed momenta are
of the form
\bea
\bp&=&\frac{2\pi}{N_c \, L}(n_1,n_2,n_3,0) + \bk \nn, \\
\bk&=&\left(\frac{\pi l_1}{L},\frac{\pi l_2}{L},\frac{\pi l_3}{L},\frac{\pi l_4}{T}\right)\;,
\eea 
where the $l_r$, for $r=1,2,3$, are integers with $-L/2 < l_r \leq L/2$
and $-T/2 < l_4 \leq T/2$ ($L$ and $T$ assumed even). The possible
entries in the integer vector $\bn = (n_1,n_2,n_3,n_4)$ depend on the 
number of directions in which the boundary condition is twisted. In our
case we have $0 \le n_1,n_2 < N_c$, $n_3=-(n_1+n_2)|_{N_c}$ and $n_4=0$. 

In NRQCD the source on the initial time slice for a Green function with 
momentum $\bp$ is
\bea
\chi(\bp,\bx)&=&\frac{ 1 }{ L^3 N_c } \, \Gamma_\bsn \,
e^{i\bsp\cdot\bsx}\;,
\nn\\
\nn\\
\Gamma_\bsn&=&z^{\half(n_1+n_2)(n_1+n_2-1)} \, \Omega_1^{-n_2} \,
\Omega_2^{n_1}\;.
\eea
We need an explicit representation for the $\Omega_\mu$, where $\mu=1,2,3$,
and for $N_c=3$ we choose
\ben
\Omega_1 = \begin{pmatrix} z&0&0\\0&1&0\\0&0&z^* \end{pmatrix}~~
\Omega_2 = \begin{pmatrix} 0&1&0\\0&0&1\\1&0&0 \end{pmatrix}~~
\Omega_3 = \Omega_1^\dagger\Omega_2^\dagger\;.
\een
In the case of purely periodic boundary conditions we can take the source for the
NRQCD Green function to be $\ONE\,\cdot\, e^{i\bsp\cdot\bsx}$, where $\ONE$
is the $N_c\times N_c$ unit matrix. This evolves all quark colour states in
one go. The analogous approach for quarks labelled by colour times smell is
not convenient and so we evolve a source appropriately chosen from the
basis of $N_c \times N_c$ matrices described above; colour and smell
singlet states, if needed, must then be constructed explicitly.
The Green function $G(\bx,\bp,t)$ satisfies the usual twisted boundary
conditions
\ben
G(\bx+L\be_\nu,\bp,t) = \Omega_\nu G(\bx,\bp,t) \Omega^\dagger_\nu\;.
\een
   
The NRQCD evolution for $G(\bx,\bp,t)$ is given by the full NRQCD
action and takes the form
\ben
G(\bx,\bp,t+1) = \sum_\bsy K(\bx,\by,t)G(\by,\bp,t)\;.
\label{eqn:greens_fn_evolution}
\een
The kernel $K$ is given by
\begin{multline}
K(\bx,\by,t) 
= \left(1-\frac{\delta H}{2}\right) \left(1 - \frac{H_0}{2n}\right)^n U_4^\dagger \\
\times \left(1 - \frac{H_0}{2n}\right)^n \left(1-\frac{\delta H}{2}\right)\;.
\end{multline}
with $H_0,\delta H$ defined in section \ref{ssec:action}.

We implement the operators in $K$ using a \verb+python+ preprocessing
package that defines each operator in $H_0$ and $\delta H$ as a
list of Wilson paths. The Wilson paths are each defined
by a list $\bs$ with a complex amplitude; these operator definitions are
read in at run time. We apply the action of each operator on
$G(\by,\bp,t)$ with a standard function that: first constructs the parallel
transporter $L(\tU;\cP(x,y\;\bs))$ for each path weighted
by the associated amplitude where $x=(\bx,t+1), y=(\by,t)$, then performs the parallel
transport of $G$ from the $t$-th to the $(t+1)$-th time slice, and finally accumulates the
results in $G(\bx,\bp,t+1)$. We solve the problem of implementing the twisted boundary
conditions in carrying out this calculation by using $\tU$ fields. The net result is that
the evolution equation can be written as
\begin{multline}
G(\bx,\bp,t+1) = \sum_m b_m\left(\sum_\bsy \cL(\tU;\cP_m) \right.\\
  \left. \times G(\by,\bp,t) R(\cP_m)\right)\;,
\end{multline}
where $\cP_m=\cP(x,y,\bs_m)$ and the sum over $m$ runs over all lists, $\bs_m$,
that define the kernel $K(\bx,\by,t)$, with $b_m$ the amplitude of the $m$-th line.
The matrix $R(\cP_m)$ implements the twisted boundary conditions and is simple to compute for
each $\cP_m$. Because $R$ right-multiplies the Green's function and
time evolution is a left-multiplying operation, we can perform the time evolution for a 
given $m$ using periodic boundary conditions for the $\tU$ fields and then independently 
right-multiply by the associated $R$ matrix. This method removes the need for any halo of 
gauge fields and the whole calculation can be easily parallelized.

Furthermore, twisted boundary conditions reduce finite-size
effects in colour singlet observables. To illustrate this result, we can
consider the example of the correlator for a meson at rest, which is given by
\ben
M(t) = \sum_{\bsy,\alpha}\Tr[G_\alpha(\by,0;t) G^\dagger_\alpha(\by,0;t)],
\een
where $\alpha$ labels the basis matrix used for the source of the quark propagator
located at the origin; all irrelevant spin degrees of freedom have been suppressed.
The correlator $M(t)$ is the sum of weighted Wilson loops consisting of a Wilson line $L_1$,
connecting $x = (0,0)$ to $y=(\by,t)$, followed by $L_2$ connecting $y$ back to $x$ and
defined by the paths $\cP_1=\cP(x,y,\bs_1)$ and $\cP_2=\cP(y,x,\bs_2)$, respectively. Then
$M(t)$ is of the form
\ben
M(t) =
\sum_{\cP_1,\cP_2,\bsy,\alpha} f(\cP_1,\cP_2)\Tr[\cL_1\chi_\alpha
R_1 R_2 \chi^\dagger_\alpha \cL_2]\;, \label{eq:Rmatrix}
\een
where $f(\cP_1,\cP_2)$ is the amplitude associated with the loop,
$(\cP_1+\cP_2)$, $\cL_i=\cL(\tU;\cP_i)$ and $R_i = R(\cP_i)$, for $i=1,2$.
Irrespective of the details of $\cL_1$ and $\cL_2$, the term sandwiched in the middle is
\ben
\sum_\alpha \chi_\alpha R_1 R_2 \chi^\dagger_\alpha = \Tr[R_1 R_2] \ONE\;.
\label{eq:Rmatrix1}
\een
Since $R_1 R_2$ is a product of the $\Omega$ matrices and their conjugates, the
trace in the above formula vanishes unless $R_1 R_2 = \ONE$. Thus, for a non-zero
contribution, the Wilson loop composed of $\cP_1$ and $\cP_2$ must loop around the spatial
torus a multiple of $N_c$ times in such a way that $R_1 R_2 = \ONE$. This reduces
finite size effects as the effective size of the lattice is now of order $N_c$ times
its spatial extent.

\subsubsection{\label{E0} Perturbative fitting of 
\texorpdfstring{$E_0$}{E0}}

We obtain the quark propagator by averaging $G(x,\bp,t)$ over
the ensemble of high-$\beta$ configurations. Because $G(\bx,\bp,t)$ is not
gauge invariant we fix the configurations to Coulomb gauge. We then
define the Coulomb ensemble-averaged quark propagator by
\ben
\hG(\bp,t,\beta,L) = \left\la \sum_{\bsx}\mbox{Re Tr}\left( \Gamma_\bsn^\dagger e^{-i\bsp\cdot\bsx}  
G(\bx,\bp,t)\right) \right\ra_{L,\beta}.
\een
Here we write $\hG(\bp,t,\beta,L)$ as a function of $L$ to indicate
explicitly that there are finite size effects, which must be accounted
for to extract the desired $L \to \infty$ result.

In order to extract the two-loop and three-loop coefficients in the perturbation expansion
for $E_0$ using the high-$\beta$ method it is necessary to carry out a
simultaneous two parameter fit in $\alpha_s$ and $L$. The
fit is a power series in $\alpha_s$ and in $1/L$ and we measure the $L \to
\infty$ coefficient of the $\alpha_s^n$, for $n=2,3$, terms.
Because the signal for the two-loop, $\alpha_s^2$, term is small compared
with the one-loop contribution the accuracy of the fit is greatly improved
by calculating the one-loop coefficient
analytically and so determining the coefficient of $\alpha_s$ in the fit.
However, Feynman perturbation theory on the lattice gives the result for lattices of large
temporal extent, $T \to \infty$, whilst here we need to carry out the perturbation 
theory for varying finite $T = 3L$. We describe the finite volume perturbation
theory for the NRQCD evolution equation in appendix \ref{sec:finite_v_pth}. It 
turns out that a minor modification of the rules for automated Feynman perturbation 
theory account for the effects of finite $T$ in the one-loop case.

%Using Eq. (\ref{eq:Grad}) and for $t$ large enough we have that
For $t$ large enough, we have that
\ben
\hG(\bp,t,\beta,L) = Z_\psi e^{(E_0+\bfp^2/2M_{pole} + \ldots)t},
\label{eq:hGfit}
\een
and by fitting to this form for a range of values of $\bp$ we can, in principle, 
extract the renormalization constants $Z_\psi, Z_{m_0}$ and $E_0$. However, for the 
current work we do not need $Z_{m_0}$ as we extract $M_{pole}$ using Eq. (\ref{eq:e0}) 
rather than Eq. (\ref{eq:zmlatt}) since, as remarked in section \ref{ssec:pmass}, the 
statistics available are not sufficient to extract a reliable value for $Z_{m_0}$. We 
therefore evaluate $\hG$ for $\bp=0$ and measure $E_0(\beta,L)$, the energy as
a function of $\beta$ and $L$.

From the boundary condition we have $\hG(\bp,t=0,\beta,L) = 1$ and so we cannot fit 
to the asymptotic form below some value $t=t_{min}$. It is a feature of Coulomb 
gauge that $Z_\psi$ is very close to unity. This is borne out by our one loop 
perturbation theory and also by simulation. Consequently, considering $Z_\psi$ and 
$E_0$ as functions of $t$, we expect the $t$ dependence of $Z_\psi$ to be small 
compared with that of $E_0$ and that $t_{min}$ is not too large . Whilst accounting 
for the need to measure in the asymptotic region by fitting only for $t \geq t_{min}$ 
it is useful to account for any residual $t$ dependence by including a transient 
function of $t$ in the exponent in Eq. (\ref{eq:hGfit}). From the finite volume 
perturbation theory and from Eqs. (\ref{eq:finiteE0}) and (\ref{eq:finiteZpsi}) 
$E_0^{(1)}(L,T,t)$ and $Z_\psi^{(1)}(L,T,t)$ depend on $t$, and a fit to their $t$ 
dependence for small $t$ gives a good indication of the explicit transient function 
we should choose. Using the one-loop calculation in this way, we find that to extract 
$E_0(\beta,L)$ from the high-$\beta$ simulation the form for $\hG$ should be chosen as
\ben 
\hG(0,t,\beta,L) = Z_\psi(\beta,L) e^{(E_0(\beta,L)t + C/t)},~~~t \geq t_{min},
\een
where, in practice, we choose $t_{min} = 5$ for all $L$. 

We fit $E_0(\beta_{pl},L)$ to a joint power series in $\alpha_V^{(n_f)}(q^*)$ and $1/L$, 
with $n_f=3$. In order to do this we need to compute the value of $\alpha_V^{(3)}(q^*)$
given the value of $\beta_{pl}$ with which the quenched configurations were generated. We first 
compute $\alpha_V^{(0)}(q^*)$ from the measured plaquette using perturbation theory. The lattice
coupling $\alpha_L$, deduced directly from the value of $\beta_{pl}$, can be expressed as a 
perturbation series in $\alpha_V^{(n_f)}(q^*)$ for any $n_f$. We eliminate $\alpha_L$ by equating 
the series for $n_f=0$ with that for $n_f=3$ and so deduce a power series for $\alpha_V^{(3)}(q^*)$ 
expanded in powers of $\alpha_V^{(0)}(q^*)$. In this way we compute the required value of 
$\alpha_V^{(3)}(q^*)$ for each value of $\beta_{pl}$. The details follow. 

We choose the $V$-scheme defined in terms of the colour Coulomb potential and the value of 
$q^*$ is found by using the BLM procedure \cite{Brodsky:1982gc,Hornbostel:2002af} applied to the 
heavy quark self-energy for determining $E_0$; M\"uller \cite{muellerthesis} gives 
$q^* = 0.794a^{-1}$ for this case. To determine $\alpha_V^{(3)}(q^*)$ given $\beta$ we use the 
value of the Wilson plaquette, $W_{11}(\beta)$, from our configurations to calculate 
$\alpha_V^{(0)}(q^*)$ using the perturbative expansion of $W_{11}$.
The BLM procedure gives the optimal value of $q^* = 3.33a^{-1}$ for this quantity 
\cite{Mason:2004aa,Mason:2005zx,Wong:2005jx}. Note that we compute 
$\alpha_V^{(0)}(q^*)$ in this manner, i.e. for $n_f=0$, since we are using quenched 
configurations. Then we have ($n_f=0$)
%\begin{widetext}
\bea
\log(W_{11}) &=& -3.068\alpha_V^{(0)}(q^*)\left(1 - 0.5945(2)\alpha_V^{(0)}(q^*)  \right.\nn \\
             && \left.  -0.589(38)\alpha_V^{(0)}(q^*)^2+ \ldots\right). 
\eea
We do not find any dependence of $W_{11}$ on $L$ since it is a short-distance, UV, 
quantity. We now relate $\alpha_V^{(0)}(q^*)$ to $\alpha_{L}(a)$ using 
\cite{Schroder:1998vy,Mason:2004aa}
\begin{eqnarray}
\alpha_{L}(a)&=&\alpha_V^{(n_f)}(q)\left(1 - v_1^{(n_f)}(q)\alpha_V^{(n_f)}(q) \right.\nn \\
        &&  \left. \ \ \ \ \ \ \ \ \ \ \ \ \ \ \  - v_2^{(n_f)}(q)\alpha_V^{(n_f)}(q)^2\right),\nonumber\\
v_1^{(n_f)}(q)&=&2\beta_0\log(\pi/q) + 3.57123 - 0.001196n_f, \nonumber\\
v_2^{(n_f)}(q)&=&2\beta_1\log(\pi/q) - [v_1^{(n_f)}]^2 + 5.382 - 1.0511n_f,\nonumber
\end{eqnarray}
%\end{widetext}
where $\beta_0$ and $\beta_1$ are the coefficients in the $\beta$-function:
\ben
\beta_0 = \frac{1}{4\pi}(11-\frac{2}{3}n_f),~~~\beta_1 = \frac{1}{(4\pi)^2}(102-\frac{38}{3}n_f), 
\een
and then use this expansion to re-express the result in terms of $\alpha_V^{(n_f)}(q^*)$.  We find
\begin{eqnarray}
\alpha_V^{(n_f)}(q)&=&
    \alpha_V^{(0)}(q)\left(1 + u_1(q)\alpha_V^{(0)}(q) + u_2(q)\alpha_V^{(0)}(q)^2\right),\nonumber\\
u_1(q)&=&v_1^{(n_f)}(q) - v_1^{(0)}(q) \nonumber,\\
u_2(q)&=&v_2^{(n_f)}(q) - v_2^{(0)}(q) + u_1(q)v_1^{(n_f)}(q). \nonumber
\end{eqnarray}
We then run $\alpha_V(q^*)$ from $q^*=3.33a^{-1}$ to $q^*=0.794a^{-1}$, appropriate 
for the fit to $E_0(\beta,L)$, using the three-loop running
\begin{eqnarray}
\frac{d\;\alpha_V(\mu)}{d\log \mu^2}&=&-\alpha_V(\mu)^2\left(\beta_0 + 
               \beta_1\alpha_V(\mu) + \beta_{2V}\alpha_V(\mu)^2 \right),\nonumber\\
\beta_{2V}&=&\frac{1}{(4\pi)^3}\left(4224.18 - 746.006n_f +20.8719n_f^2\right),\nonumber
\end{eqnarray}
where we suppress the $n_f$ superscript from now on, using $n_f=3$ implicitly.

\begin{figure}
\includegraphics[width=0.98\hsize]{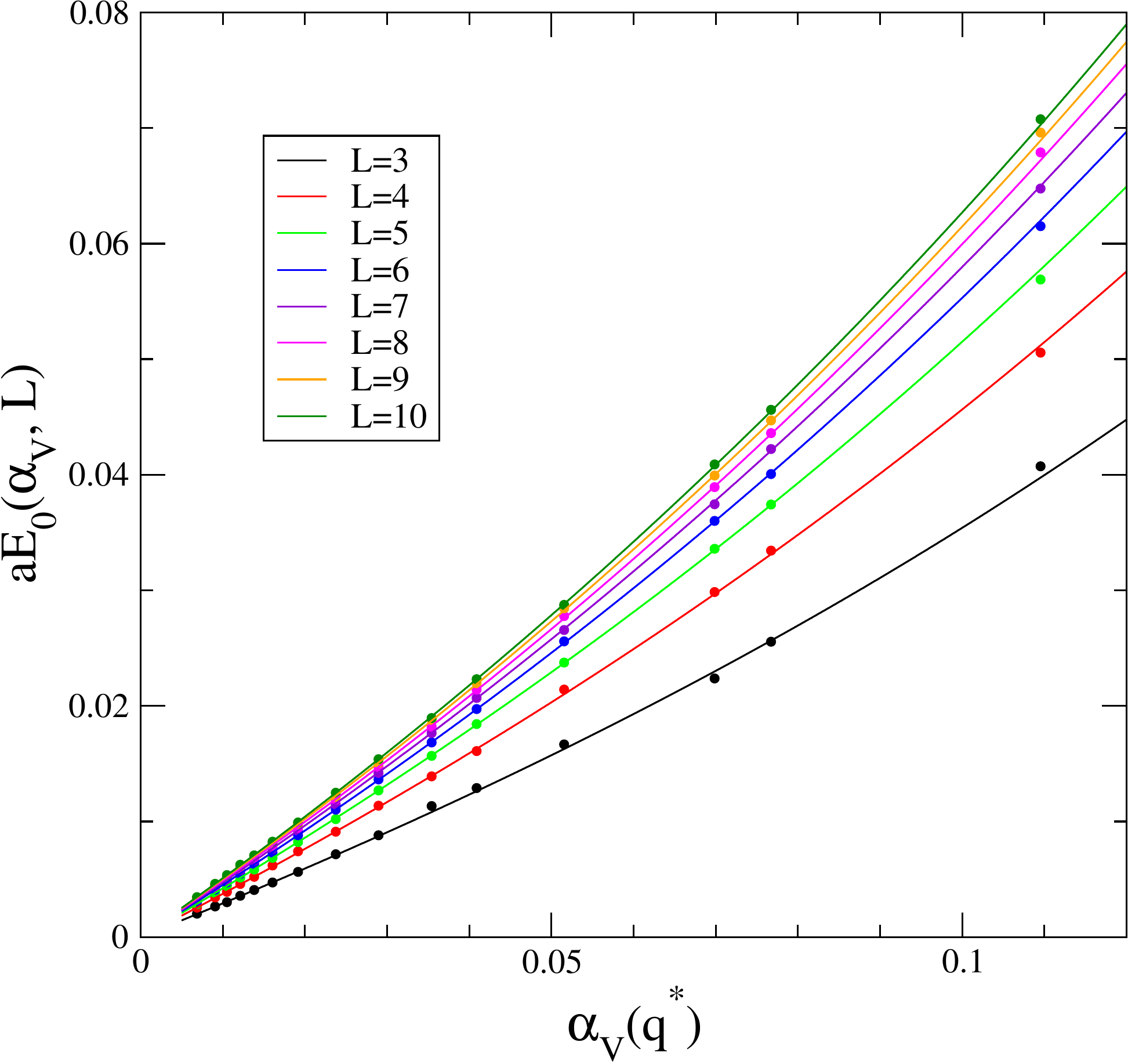}
\caption{\label{fig:E0_1.72}$E_0(\alpha_V(q^*),L)$ for $aM=1.72$ for both data and fit for the
values of lattice size $L^3 \times T$, $T=3L$ used in the extraction of the two-
and three-loop quenched coefficients in the perturbation series for $E_0$. 
We write $E_0$ as a function of the $\alpha_V(q^*)$ value rather than $\beta_{pl}$.
Here, $q^* = 0.794 a^{-1}$. This fit has $\chi^2 = 1.2$.} 
\end{figure}

We fit $\hG(0,t,\beta,L)$ separately, as discussed above, for the set of $\beta,L$ 
values and deduce $E_0(\beta,L)$.
% using a binning procedure to determine the statistical errors. 
As the data may contain residual auto-correlations, we resample via
blocking to determine the true statistical error.  Within independent 
chains, sequential measurements are grouped together into bins and 
the means of each bin are treated as statistically independent. The size of 
the bins is determined by examining the scaling of the variance as 
a function of the bin size, and is dependant on the values of 
$ L $ and $ \beta_{ pl } $, and the operator being measured.
We then fit these values to the form
\begin{multline}
E_0(\beta,L) = (E_0^{(1)}(L,T/2)+\delta)\alpha_V(q^*)  \\
+ (c_{20}+\frac{1}{L}c_{21})\alpha_V(q^*)^2       
%&&
+c_{30}\alpha_V(q^*)^3,
\end{multline}
with $q^* = 0.794a^{-1}$ and $T=3L$.
 Here $E_0^{(1)}(L,T/2)$ is 
the calculated value for the one-loop contribution which includes the contribution 
from tadpole improvement of the NRQCD Hamiltonian; this contribution is a constant, independent of 
$\beta$ and $L$. We allow for a small additive adjustment $\delta$, independent of $\beta_{pl}$ and $L$,
in the values of the $E_0^{(1)}(L,T/2)$  accounting for any minor mismatch between their analytical 
and numerical calculation; as we should expect, $\delta$ is found to be very small. The finite-size, $L$, 
dependence of $E_0$ is included in $E_0^{(1)}(L,T/2)$ and in the two-loop 
coefficient. We find that this parametrization is sufficient for a very good fit to 
the data; within errors we do not discern any $\alpha^2/L^2$ or $\alpha^3/L$ 
contributions. The fit is for 116 degrees of freedom (4 parameters, 15 $\beta$ values 
and 8 $L$ values) and we find $\chi^2 = 1.2$ and  $1.1$, respectively, for $am_0= 1.72, 
2.5$. In Fig \ref{fig:E0_1.72} we show $E_0(\beta,L)$ plotted versus $\alpha_V(q^*)$ for the 
different $L$ and for $am_0=1.72$. The quenched results that we require are 
$E_0^{(2),q} = c_{20}$.

\section{\label{sec:nonpert}Nonperturbative determination of \texorpdfstring{$E_{\rm sim}$}{Esim}}

We now briefly discuss the nonperturbative determination of the meson energies $E_{\rm sim}$. The method is standard and this NRQCD action \cite{dowdall12} has been thoroughly tested by HPQCD in a range of calculations.

We use two ensembles of gauge configurations generated by the MILC collaboration with $n_f=2+1$ ASQtad sea quarks, which we denote coarse ($\sim$0.12 fm) and fine ($\sim$0.09 fm) \cite{Aubin:2004wf,Bazavov:2009bb}. Details are given in table \ref{tab:gaugeparams}.
The light quark masses on these ensembles are not particularly chiral but we have seen that the light sea quark mass has negligible effect on most quantities in the bottomonium spectrum \cite{dowdall12}.
The lattice spacing on these ensembles has been determined using the static quark potential parameter $r_1$ in \cite{Davies:2009tsa}, and 
is given in the table.

\begin{table}
\caption{
Details of the two ASQtad gauge configurations used in the nonperturbative determination of $E_{\rm sim}$.
$\beta$ is the gauge coupling, $a^{-1}$ is the inverse lattice spacing determined using the static quark potential parameter $r_1$,
$u_0am_{l}$, $u_0am_{s}$ are the light sea quark masses, $L$ and $T$ are the lattice dimensions and $n_{{\rm cfg}}$ the size of the ensemble.
}
\label{tab:gaugeparams}
\begin{ruledtabular}
\begin{tabular}{llllllll}
Set & $\beta$ & $a^{-1}$ (GeV)  & $u_0am_{l}$ & $u_0am_{s}$ & $L \times T$ & $n_{{\rm cfg}}$  \\
\hline
coarse & 6.76 & 1.652(14)  & 0.01   & 0.05   & 20$\times$64 & 1380 \\
fine   & 7.09 & 2.330(17)  & 0.0062 & 0.0310 & 28$\times$96 & 904 \\
\end{tabular}
\end{ruledtabular}
\end{table}

The NRQCD action is given in section \ref{ssec:action} and includes one-loop radiative corrections to the coefficients calculated in \cite{dowdall12,Hammant:2011bt}. 
The coefficients are listed in table \ref{tab:ciparams}. 
The same coefficients are used in the perturbative calculations and in the high-$\beta$ simulations, but with $\alpha_s$ evaluated at a scale appropriate for $\beta$, as discussed in section \ref{ssec:action}.

\begin{table}
\caption{
Coefficients used in the nonperturbative simulation.
$u_{0,P}$ is the plaquette tadpole improvement factor, $c_i$ are the coefficients in $\delta H$.
}
\label{tab:ciparams}
\begin{ruledtabular}
\begin{tabular}{llllllllll}
Set & $u_{0,P}$ & $c_1$ & $c_2$ & $c_3$ & $c_4$ & $c_5$ & $c_6$ &  \\
\hline
coarse & 0.86879  & 1.31 & 1.0 & 1.0 & 1.2  & 1.16 & 1.31\\
fine   & 0.878214 & 1.21 & 1.0 & 1.0 & 1.16 & 1.12 & 1.21
\end{tabular}
\end{ruledtabular}
\end{table}

Tuning the bare $b$-quark mass accurately is an important part of the calculation as this is a potential source of error in $\msms$. The heavy quark energy shift means that we cannot tune using the meson energy directly but we must use the kinetic mass determined from the dispersion relation, which is much noisier. A detailed study of the systematic errors incurred and their effect on the accuracy of the bare mass was carried out in Ref \cite{dowdall12}.
To reduce systematic errors we use the spin average of the vector and pseudoscalar bottomonium states
\ben
\overline{M_{b \bar b}}= ( 3M_{{\rm kin},\Upsilon} + M_{{\rm kin},\eta_b} )/4,
\een
which eliminates errors from missing spin dependent higher order terms and radiative corrections in the action.
We must also take account of missing electromagnetic effects, sea charm quarks and annihilation of the $\eta_b$ to gluons by shifting the experimental values appropriately. These effects were estimated in \cite{Gregory:2010gm}, resulting in an adjusted experimental value of $M_{b\bar b}^{\rm expt} = 9.450(4)$ GeV, where the error comes from taking a large uncertainty on the shifts that were applied.
The correctly tuned bare $b$-quark masses in lattice units that we obtain are $2.49(2)_{\rm stat}(1)_{\rm sys}$ on the coarse lattice, and 
$1.71(2)_{\rm stat}(1)_{\rm sys}$. The first error includes a sizeable statistical error from the kinetic mass and all lattice spacing errors, the second includes the systematic errors in the kinetic mass estimated in \cite{dowdall12}. The effect of these errors are included in the final error budget.

The valence strange quark propagators used in the $B_s$ mesons use the Highly Improved Staggered Quark (HISQ) action \cite{Follana:2006rc} and are tuned using the $\eta_s$ meson. 
This is a fictitious $s \bar s$ particle which, with the addition of experimental data for $M_\pi,M_K$ and chiral perturbation theory, is a very convenient choice for tuning the $s$ mass and fixing the scale. 
The value on the $n_f=2+1$ ensembles that we are using is $M_{\eta_s}=0.6858(40)$ GeV \cite{Davies:2009tsa}.

The ground state energies $E_{\rm sim}$ are extracted from multiexponential Bayesian fits \cite{gplbayes} to meson correlation functions that use multiple smeared sources for the 
quark propagators. To further improve statistics we used stochastic noise sources and ran 16 time sources on each configuration for the $\Upsilon$, and 4 for the $B_s$. The results are listed in table \ref{tab:E0_simres}.

\begin{table}
\caption{\label{tab:E0_simres}
Masses and extracted energies from the nonperturbative simulations.
$am_0$ and $am_s$ are the bare (valence) $b$ and $s$ masses, $aE_{\mathrm{sim},X}$ are the fitted ground state energies of the meson X in lattice units. The first row is for the coarse ensemble and the second for fine.
The errors are from statistics/fitting only.}
\begin{ruledtabular}
\begin{tabular}{cccccc}
$am_0$ & $am_s^{\rm val}$ &$aE_{\mathrm{sim},\Upsilon}$ & $aE_{\mathrm{sim},\eta_b}$ 
& $aE_{\mathrm{sim},B_s}$ & $aE_{\mathrm{sim},B_s^\ast}$ \\
\vspace*{-10pt}\\
\hline
\vspace*{-8pt}\\
2.50 & 0.0496 & 0.46591(6) & 0.42579(3) & 0.6278(5) & 0.6595(6) \\
1.72 & 0.0337 & 0.41385(4) & 0.38124(2) & 0.4812(5) & 0.5027(7) \\
\end{tabular}
\end{ruledtabular}
\end{table}

\section{\label{sec:mb-msbar}Calculating the \texorpdfstring{$\MS$ $b$-quark}{MS bar b quark} mass}

Now that $E_0$ and $E_{\rm sim}$ have been determined, we can combine the results into a perturbative series for $\msms$ in the $\MS$ scheme. This requires various scheme conversions and changes of scale to give the series at the scale relevant for the $b$-quark mass.
This then gives the result at $n_f=3$ and we can use known formulas to convert this to the usual $n_f=5$ result.
We repeat this whole process at both values of the bare mass to check for discretisation errors which will then be included in our error.

To further reduce systematic errors, we adjust Eq \ref{eq:e0} so that we use the spin-averaged bottomonium mass $M_{b\bar b} = (3M_\Upsilon + M_{\eta_b})/4 $. This removes any error from spin dependent terms in the NRQCD action. As discussed in section \ref{sec:nonpert}, the experimental result 
used must be adjusted to $M_{b\bar b, \rm expmt} = 9.450(4)$ GeV to 
reflect the absence of electromagnetism, 
sea charm quarks and $\eta_b$ annihilation.

\subsection{Perturbative series for \texorpdfstring{$\msms$}{MS bar mass}}

So far, all our perturbative results have been expressed in terms of
$\alpha_V$, the coupling constant defined in the $V$-scheme at the scale $q^*=0.794/a$. 
\begin{align}\label{eq:E0V}
aE_0 = {} & aE_0^{(1)}\alpha_{V}(q^*)\nn + \left(aE_0^{(2)} + 
               aE_0^{u0,f} \right)\alpha_{V}^2(q^*)  \nn \\
      {} & + aE_0^{(3),q}\alpha_{V}^3(q^*).
\end{align}
The results for each component are given in table \ref{tab:E0_pertres}.

%The 2-loop $n_f$ contribution was evaluated at a different scale $\hat{q}=3.33/a$, running this to $q*$ with 
%the beta function gives a small 3-loop $n_f$ contribution $aE_0^{(2\hat{q})}$ not included in the above formula.
The series expansion of $aE_0$ is truncated at $\alpha_s^3$ and we take $n_f=3$ as this is the number of sea quarks in the nonperturbative determination of $E_{\rm sim}$. No fermionic $\alpha_s^3$ contributions are included in the series.
The effects of the one loop tadpole corrections
are directly included in the tadpole improved results from the high-$\beta$
simulation, as are the quenched two loop tadpoles. However,
the two loop fermionic tadpole contributions are not included in the high-$\beta$
results so we must add the corresponding correction, $aE_0^{u0,f}$, to the energy shift.
$aE_0^{u0,f}$ is given by
\cite{gulez04}
\begin{align}\label{eq:e0ftad}
aE_0^{u0,f} = {} & \left[1+\frac{7}{2am_0} -
\frac{3}{2}\left(\frac{1}{a^3m_0^3}+\frac{1}{2na^2m_0^2}\right)\right]u_0^{(2),f},
\end{align}
where $u_0^{(2),f}$ is the fermionic contribution to $u_{0,P}$ given in section \ref{sec:auto-lpt}.

The other perturbative factor that we need is the pole to $\MS$ renormalization $Z_M$
which is reproduced in appendix \ref{sec:matching_formulas}.
Inserting these two series into eq. \eqref{eq:msbar} gives a series for $\msms$.

We now relate $\alpha_V(q^{\star})$ to $\alpha_{\MS}(q^{\star})$. This is done using the 
three-loop relation in \cite{peter97a,peter97b,schroeder99} which is summarised in appendix 
\ref{sec:matching_formulas}, and express $E_0$ as a series in the $\MS$ scheme.
Matching is done at $q^*$ to avoid logarithmic contributions. 
The series is then run to $\mu=4.2$ GeV using the 4-loop $\MS$ beta function.

To evaluate the series we need the relevant value of $\alpha_{\MS}$, which in this case is the 3-flavour value at $\mb$. 
Since $\MS$ is a mass independent scheme, high mass particles do not explicitly decouple from the beta function and one must construct an effective theory with $n_l=n_f-1$ quarks when crossing a quark mass threshold \cite{Prosperi:2006hx}. This introduces discontinuities in the running of $\alphams$ at the thresholds 
which have been calculated to 4-loops in \cite{Chetyrkin:1997sg}, and we give the relevant formulas in appendix  \ref{sec:matching_formulas}.
We start with the current PDG average $\alpha_{\MS}(M_Z,n_f=5)=0.1184(7)$ which we run to 4.2 GeV using the 4-loop running with $n_f=5$ \cite{vanRitbergen:1997va}, then matching to the $n_f=4$ theory and running down to $1.2$ GeV to match to $n_f=3$, before running back up to 4.2 GeV with $n_f=3$ running. We find $\alpha_\MS(\mb,n_f=3)=0.2159(20)$. Small changes in the matching scales have negligible effect on the value.

Using this value of the coupling the results using $M_{\bar b b} $ are $\mbmbnf = 4.195(8)$ GeV on the coarse lattice and 
$\mbmbnf = 4.198(10)$ GeV on the fine lattice.
We also tried allowing the scale to float and solving such that $\mu$ was exactly the $\MS$ mass but this
makes negligible difference to the result.
The results using the $B_s$ mass give $\mbmbnf = 4.177(8)$ GeV on the coarse lattice and 
$\mbmbnf = 4.191(10)$ GeV on the fine lattice. These are consistent with the bottomonium results.
This error includes statistical errors in the perturbation theory integrals, lattice spacing error, and simulation errors in the ground state masses (negligible). We have not yet included an estimate of the truncation error in the perturbative series.

Our calculations were performed using lattice results with $n_f=3$ sea quarks. In order to compare to the real world we must match this value to $n_f=5$. As with the coupling constant, a running quark mass in a mass independent scheme is discontinuous at flavour thresholds and must be matched to an effective theory with a different number of flavours. The formula for the mass decoupling is given in appendix \ref{sec:matching_formulas} in equation \eqref{mdec}. We run down to 1.2 GeV with three flavour mass running \cite{vanRitbergen:1997va},  match to a theory with $n_f=4$, run up to 4.2 GeV and match to the $n_f=5$ theory. Again, small changes to the matching scale or the final scale at which we evaluate the mass have negligible effect. 
After this running, the values we obtain for the $M_{\bar b b}$ results are $\mb\left(\mb,n_f=5\right) = 4.161(10)$ GeV on the coarse lattice and 
$\mb\left(\mb,n_f=5\right) = 4.164(12)$ GeV on the fine lattice where from now on we state $n_f$ explicitly.
Overall, matching to the $n_f=5$ theory shifts the mass down by around 30 MeV.

In principle there may be discretisation errors arising from lattice artefacts. Since we have two lattice spacings available we can fit the results as a function of $a$ to obtain the physical result and to allow a systematic error for this dependence. In fact the dependence is very mild as is clear from the fact that all 
of the results are consistent with each other. Our NRQCD action contains discretisation corrections that get renormalized as a function of the cutoff $am_0$ and so we allow an additional mild dependence of the fit function on $am_0$. This makes no difference to the fit. The form is
\begin{multline}
\label{eq:fitbsextrap}
\mbmb(a,\delta x_m) =
\mbmb
 \\
\times \left[ 
1  
+  \sum_{j=1}^2 d_j(\Lambda a)^{2j}(1 + d_{jb}\delta x_m + d_{jbb}(\delta x_m)^2)
\right], 
\end{multline} 
where we have allowed discretisation effects with a scale of $\Lambda=0.5$ GeV
and cutoff dependence via $\delta x_m = (am_0-2.1)/(2.5-1.7)$ which varies between $\pm0.5$.
Priors on the values are 4.2(5) for the mass, 0.0(3) for the $a^2$ term since our action is one-loop improved, and 0(1) for everything else.

Some of the errors in the data are correlated and we allow for this in the fit. 
We multiply the $\mb$ values by a $(1+n_f\alpha_s^3)$ truncation error (discussed below) which is 100\% correlated between the points on the two lattice spacings.
The errors on all quantities coming from the high-$\beta$ simulations are correlated with corresponding errors on the other lattice spacing. Statistical errors coming from \vegas integrals are uncorrelated.

We only fit the bottomonium results as the $B_s$ results are in very good agreement. The result of the fit is
$
\mb\left(\mb,n_f=5\right) = 4.166(42){\rm \ GeV}.
$

\begin{table}
\caption{\label{tab:E0_pertres}Perturbative results required to extract the $\MS$ mass. 
The quenched results, indicated by superscript $q$, are from high-$\beta$ simulations. 
The one-loop data are the exact perturbative results extrapolated
to infinite lattice size. The two loop results include both quenched and
fermionic contributions. The three-loop values include only quenched
results. We evaluate all results in the $V$-scheme at a characteristic scale 
of $q^{\star} = 0.794a^{-1}$ \cite{muellerthesis}.%\cite{Muller:2009aa}.
}
\begin{ruledtabular}
\begin{tabular}{ccccc}
$am_0$ & $aE_0^{(1)}$ & $aE_0^{(2)}$ & $aE_0^{u0,f}$ & $aE_0^{(3),q}$ \\
\vspace*{-10pt}\\
\hline
\vspace*{-8pt}\\
2.50 & $0.6786(1)$ & $1.16(4) - 0.2823(6)n_f$ & $0.158531(16)n_f$ & $2.3(3)$ 
\\
1.72 & $0.5752(1)$ & $1.30(4) - 0.3041(3)n_f$ & $0.186607(19)n_f$ & $2.3(3)$
\\
\end{tabular}
\end{ruledtabular}
\end{table}

\begin{figure}
 \begin{center}
  \includegraphics[width=0.99\hsize]{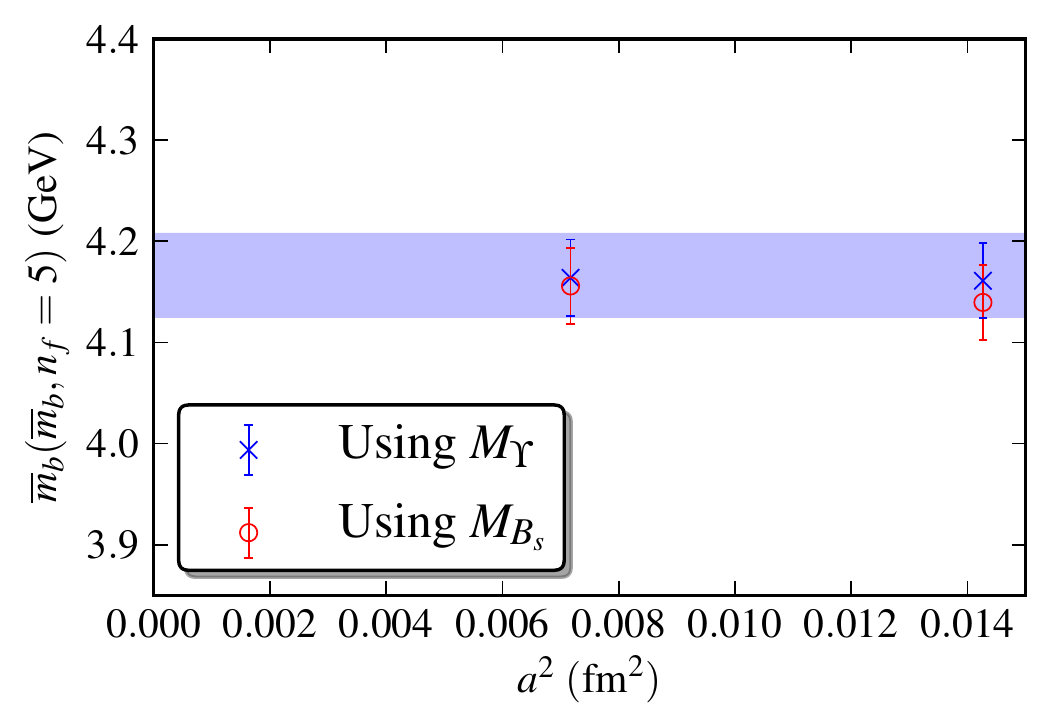}
 \end{center}
\caption{\label{plot:mb} Results for the $n_f=5$ $\MS$ mass using both bottomonium and $B_s$ meson simulation data, and the fit to the bottomonium results. The errors on the data points include statistics, error on $\alpha_{\MS}$ and a correlated truncation error on the perturbative series. Additional (subdominant) errors are described in the text.}
\end{figure}

\subsection{Error budget}\label{ssec:errors}

Broadly, the  three main sources of uncertainty in our result for the
$b$-quark mass are: statistical errors, errors from truncating the perturbation series and other systematic errors. 
We expect the ${\cal O}(\alpha_s^3)$ perturbative contributions to dominate
the uncertainty in our final result. In this section we
discuss each of these sources of
error in turn and tabulate our error budget in Table
\ref{tab:errorbudget}.

\paragraph{Statistical errors} Statistical errors arise in the
nonperturbative calculation of $E_{\mathrm{sim}}$, and in the contributions at each order in
the expansion of the heavy quark energy shift, $E_0$. 
The statistical error in $E_{\mathrm{sim}}$ comes from the fit to lattice 2-point functions and is completely negligible.
The statistical error in the one-loop piece of $E_0$ comes from the evaluation of diagrams using \vegas and from the extrapolation to infinite volume.
The uncertainties in the two loop and three-loop quenched coefficients of $E_0$ arise from the simultaneous fit
to $\alpha$ and $L$. This is significant at 14 MeV. 
The statistical error in the two loop fermionic coefficient is due to the numerical evaluation of
the Feynman diagrams and the extrapolation to zero light quark mass.

\paragraph{Perturbative errors}

The three-loop fermionic contribution to the energy shift is unknown, so we
estimate the error due to this contribution as ${\cal O}(n_f \times
\alpha_{\MS}^3)$. This is the dominant source of error in our calculation.
Perturbative errors from running the coupling and quark mass are negligible as the formulas are higher order.

The fermionic contributions are the only unknown source of
uncertainty at three-loops in our result. In principle these effects can be
calculated using automated lattice perturbation theory. However, there are
a large number of diagrams to evaluate, many of which are likely to have
complicated pole structures and possible divergences (the energy shift is
infrared finite, but individual diagrams may have divergences that
ultimately cancel). The complexity of such a calculation would be
considerable.

\paragraph{Other systematic errors} 
\begin{itemize}
 \item Bare mass tuning: The tuning of the bare $b$-quark mass used in $E_0$ and $E_{\rm sim}$ is a source of error. We can estimate the error due to mistuning using the errors given on the tuned masses $2.49(2)_{\rm stat}(1)_{\rm sys}$, and $1.71(2)_{\rm stat}(1)_{\rm sys}$ and by estimating the bare mass dependence of each quantity. We use only the one-loop piece of $E_0$ and compute the value at an extra mass, we find a linear dependence with a slope of 0.13. For $E_{\rm sim}$, we use the results at different bare masses given in \cite{dowdall12} and find a dependence that is less than 0.01, which we take to be linear for these small increments. By recomputing $\msms$ taking a $1\sigma$ deviation in the bare mass, we find errors of 4 MeV on the coarse lattice, and 6 MeV on fine. We take the larger of these as an error on our result.

 \item Corrections for missing electromagnetism, charm quarks in the sea and $\eta_b$ annihilation were estimated and applied to the experimental $\Upsilon,\eta_b$ masses. We add the errors linearly rather than in quadrature and propagate this error through to the final result, which gives 1.9 MeV. 

 \item Higher order relativistic corrections: These arise from not including  $\mathcal{O}(v^6)$ terms in our NRQCD action and, with $v^2\sim0.1$, could contribute 1$\%$ of the binding energy which is 5 MeV. 
 \item Radiative corrections: $\alpha_s^2 v^4$ should be smaller at around half a percent of the binding energy so we take 2.5 MeV.
 \item Lattice spacing errors, including $r_1/a$: These are included as the ``statistical'' error on the data points in the plot but we estimate their contribution to the final error to be 4.5 MeV.
 \item Lattice spacing dependence: We incur an error from fitting the two masses as a function of $a$ which we can estimate from the fit. The lattice spacing dependence is not significant but we find 16 MeV, this is already included in the total error quoted from the fit.
 \item Sea quark mass dependence: We have only used one sea quark mass in our calculation but in previous calculations we have observed very mild dependence in $E_{\rm sim}$ \cite{dowdall12}.
Errors from light sea quark mass dependence should be negligible compared to our other errors.
\end{itemize}

With these errors included, our final result for the $\MS$ $b$-quark mass is:
\begin{equation}\label{eq:mbaverage}
\mb(\mb,n_f=5) = 4.166(43)
\,\text{GeV}.
\end{equation}

\begin{table}
\caption{\label{tab:errorbudget}The $b$-quark mass error budget,
systematic error estimates are discussed in more detail in the text.
}
\begin{ruledtabular}
\begin{tabular}{lcc}
Source &  Error (MeV) & Error (\%) \\
\vspace*{-10pt}\\
\hline
 $n_f\alpha_s^3$ perturbative error     & 36    & 0.9   \\ 
 $M_{\Upsilon},M_{\eta_b}$ experiment 	& $< 0.1$ & $< 0.01$ \\
 $aE_{\mathrm{sim}}$ 			& $< 0.1$ & $< 0.01$	\\
 $am_0$ tuning 				& 6       & 0.14 \\
 \vegas integration			& $< 0.1$ & $< 0.01$ \\
 High-$\beta$ statistics		& 14      & 0.35  \\ 
 $a$ dependence 	& 16 	& 0.38  \\
 Scale uncertainty 	& 4.4 	& 0.10  \\
 $\alpha_s$ uncertainty & 0.2 	& 0.01  \\
 Relativistic $v^6$ 	& 5 	& 0.12  \\
 Radiative $\alpha_sv^4$& 2.5 	& 0.06  \\
 E\&M, Charm sea, annih.& 1.9 	& 0.05  \\
\vspace*{-10pt}\\
\hline
\vspace*{-8pt}\\
Total   & 43 MeV & 1.0 \%  \\
\end{tabular}
\end{ruledtabular}
\end{table}

\section{\label{sec:discussion}Discussion}

We can compare our result to previous values from 
the literature. As discussed in Section \ref{sec:intro}
there are a number of accurate theory results from 
comparing continuum QCD perturbation (through $\alpha_s^3$) 
for moments of the vector charmonium current-current correlator
to experimental results extracted from 
$\sigma(e^+e^- \rightarrow \mathrm{hadrons})$ in the 
$b$ region. In \cite{Chetyrkin:2009fv}, for example, the result 
$\overline{m}_b(\overline{m}_b) = 4.163(16)\,\mathrm{GeV}$ 
is obtained. In \cite{mcneile10} lattice QCD calculations of 
time-moments of the $\eta_b$ correlator are used 
instead of the experimental results to give  
$\overline{m}_b(\overline{m}_b) = 4.164(23)\,\mathrm{GeV}$. 
It was important in this calculation to use 
pseudoscalar correlators in a lattice QCD 
formalism (HISQ) that has absolutely normalised 
pseudoscalar currents. Our result agrees with these 
two values. It is not as accurate because we are not 
using such high order QCD perturbation theory but 
it nevertheless provides a check from a completely 
different perspective at the level of 1\%. 

There are also a number of results using alternative 
methods from lattice QCD but these are not typically 
very accurate. An early result for $\overline{m}_b$ with NRQCD $b$-quarks 
on the $n_f=2+1$ MILC configurations including $u$, $d$ 
and $s$ sea quarks was 4.4(3) GeV \cite{gray05}, the 
large error here arising from the use of one-loop 
lattice QCD perturbation theory for $Z_M$.  
More recently, methods have been developed by the 
ALPHA collaboration for 
determining the energy 
shift for lattice Heavy Quark 
Effective Theory nonperturbatively, including next-to-leading-order 
terms in the inverse heavy quark mass expansion for 
the valence $b$-quarks \cite{Bernardoni:2012fd}. This has been 
implemented on gluon field configurations including 
$u$ and $d$ sea quarks in the clover formalism. Combining with the 
experimental $B$ meson mass in a similar approach to the one used 
here, gives  
$\overline{m}_b(\overline{m}_b) = 4.22(11)\,\mathrm{GeV}$. 
The error here is dominated by lattice statistical and systematic errors. 
Another method by the ETM collaboration \cite{Dimopoulos:2011gx} uses a ratio of quark masses to 
heavy-light meson masses with a known infinite mass limit. 
This is implemented on gluon field configurations including 
$u$ and $d$ sea quarks in the twisted mass formalism and 
valence $b$ and light twisted mass quarks. 
Interpolating to the $b$-quark and using experimental meson 
masses gives: 
$\overline{m}_b(\overline{m}_b) = 4.29(14)\,\mathrm{GeV}$, 
with an error dominated by lattice statistical errors. 
Note that neither of the ALPHA or ETM results include $s$ quarks in 
the sea and the error from this is not estimated. 

\begin{figure}
 \begin{center}
  \includegraphics[width=0.99\hsize]{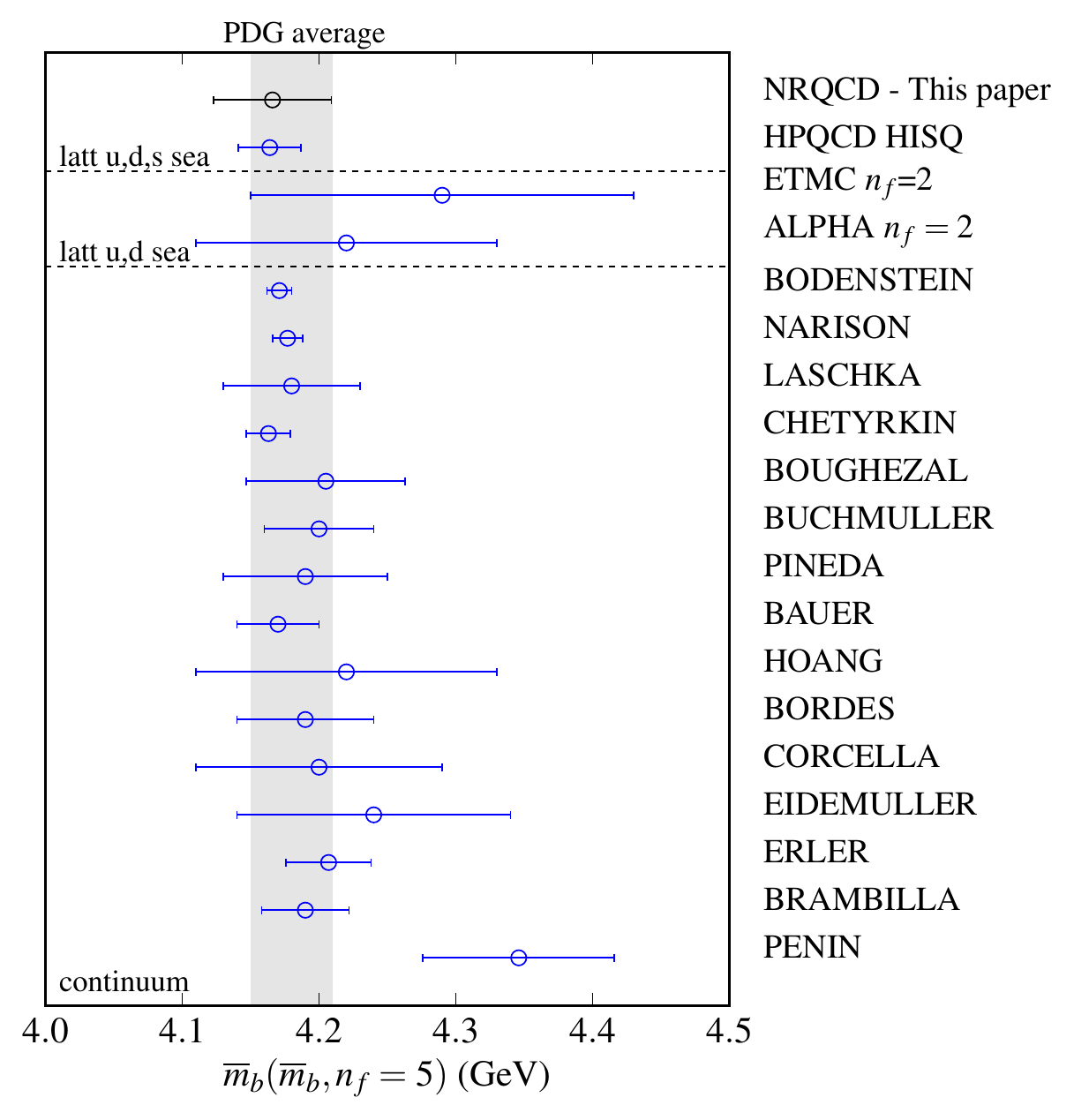}
 \end{center}
\caption{\label{plot:comparison} Comparison of our result with other recent theory-based $b$-quark mass determinations. We include all determinations listed in the PDG summary table \cite{PDG:2012} but separate lattice QCD determinations with $n_f=2$ and $n_f=2+1$ sea quarks for easier comparison 
\cite{mcneile10,Dimopoulos:2011gx,Bernardoni:2012fd,Bodenstein:2011fv,Chetyrkin:2009fv,Narison:2011rn,Laschka:2011zr,Boughezal:2006px,Buchmuller:2005zv,Pineda:2006gx,Bauer:2004ve,Hoang:2004xm,Bordes:2002ng,Corcella:2002uu,Eidemuller:2002wk,Erler:2002bu,Brambilla:2001qk,Penin:2002zv}. 
}
\end{figure}

Fig.~\ref{plot:comparison} collects a number of lattice and continuum 
QCD determinations of the $b$-quark mass for comparison. 
The evaluation of 4.18(3) GeV in the Particle 
Data Tables~\cite{PDG:2012} is shown by the grey band. 
There is good consistency between all determinations including 
the new result of this paper.

\section{\label{sec:conclusion} Conclusion}

In this paper we have presented a new determination of the $b$-quark mass
from simulations of lattice NRQCD at two heavy quark masses. The
uncertainty associated with previous determinations of the $b$-quark mass
from lattice NRQCD was dominated by the one loop perturbative calculations
used to extract the $b$-quark mass. By calculating the heavy quark energy
shift at two loops, we have significantly reduced this uncertainty. The resulting error is now in line with the
most precise lattice determinations available.

In order to efficiently calculate renormalisation parameters at two
loops, we implemented a mixed approach, combining quenched high-$\beta$
simulation with automated lattice perturbation theory. We were also able
to extract estimates of the three loop quenched contributions to the
energy shift from high-$\beta$ simulations and found that all perturbative
coefficients are well-behaved. The reliable extraction of the two loop
energy shift convincingly demonstrates the effectiveness of our approach.

As part of this calculation, we also determined the fermionic
contributions to the two loop tadpole improvement factor for both the
Landau and plaquette tadpole definitions.

We undertook a number of checks of both the automated lattice perturbation
theory and the high-$\beta$ simulations. For the former, we confirmed that
we could reproduce published one loop results, that the energy shift was
infrared finite and that the fermionic insertions in the gluon propagator
obeyed the relevant Ward identity. For the latter, we were able to compare
one loop results to the exact finite size perturbation theory results to
ensure the correctness of our fits.

The uncertainty in our result is now dominated by the unknown fermionic
contributions to the three loop energy shift, which is in principle
calculable with automated lattice perturbation theory. Greater statistics in the
high-$\beta$ simulations may also allow us to extract the quenched
contributions to the mass renormalisation with sufficient precision to
enable an independent determination of the $b$-quark mass by direct
matching.

%%%%%%%%%%%%%%%%%%%%%%%%%%%%%%%%%%%%%%%%%%%%%%%%%%%%%%%%%%%%%%%%%%%%%%%%%%%%%%%%
% Acknowledgments
%%%%%%%%%%%%%%%%%%%%%%%%%%%%%%%%%%%%%%%%%%%%%%%%%%%%%%%%%%%%%%%%%%%%%%%%%%%%%%%%
%%%%%%%%%%%%%%%%%%%%%%%%%%%%%%%%%%%%%%%%%%%%%%%%%%%%%%%%%%%%%%%%%%%%%%%%%%%%%%%%
${}$%\\
\acknowledgments
We thank Matthew Wingate and Laurent Storoni for useful discussions.
We thank the DEISA Consortium,
%(\url{http://www.deisa.eu}),
co-funded through the EU~FP6 project RI-031513 and the FP7 project RI-222919,
for support within the DEISA Extreme Computing Initiative. 
This work was supported by STFC under grants ST/G000581/1 and ST/H008861/1.
The calculations for this work were, in part, performed on the University
of Cambridge HPCs as a component of the DiRAC facility jointly funded by
STFC and the Large Facilities Capital Fund of BIS.
CJM is supported by DoE grant DE-FG02-04ER41302.
CTHD is supported by the Royal Society and the Wolfson Foundation.

\appendix

%\section{\label{sec:renorm_consts} Renormalization constants}
%\input{mb_renorm_const.tex}

\section{\label{sec:finite_v_pth} Finite volume perturbation theory}

Without loss of generality we consider a scalar model that is  sufficient to demonstrate the
approach. 
We take the NRQCD evolution for the heavy quark Green function to
be
\ben
\tG(\bp,t) = \tK(\bp,t-1)\tG(\bp,t-1),
\een
where 
\bea
\tK(\bp,t)&=& \tK_0(\bp,t)\left(1-\frac{g}{2}\phi\right),\nn\\
\tK_0(\bp)&=&\left(1-\frac{p^2}{2mn}\right)^n.
\eea
Here $K(\bx,t)$ is the approximation to the evolution operator $e^{-H}$
with 
\ben
H = \left(1+\frac{\nabla^2}{2m}-g\phi\right). 
\een
We then have that
\ben
\tG_0(\bp,t) = \tK(\bp)^t,~~\mbox{with}~~\tG_0(\bp,0) = 1.
\label{G0}
\een
The diagram we consider is the rainbow diagram.
The vertices are labelled with $(\bp,t)$ coordinates, appropriate to the Hamiltonian formalism.
The vertices are separated by time $\tau$. The rainbow diagram has $\tau > 0$ whilst
the associated tadpole diagram has $\tau = 0$. There is no effect of finite $T$ on the
calculation of the tadpole diagram, which is therefore given by finite-$L$
Feynman perturbation theory. At $O(g^2)$ from the diagram we have the contribution
\ben
\tG_2(\bp,t) = g^2\sum_{\bsq,\tau=1}^{T-1} (t-\tau)\tK_0(\bp)^{t-\tau}\Gamma(\bq,\tau)\tK_0(\bp-\bq)^\tau.
\een
The factor $(t-\tau)$ is the number of temporal positions the graph can adopt and
$\Gamma(\bq,\tau)$ is the 
$\phi$-field propagator, given by
\bea
\Gamma(\bq,\tau)&=&\frac{1}{T}\sum_{\Omega=0}^{T-1} \tGamma(\bq,q_0)
e^{iq_0\tau},\nn\\
\tGamma(\bq,q_0)&=&\frac{1}{\hq^2+\hqzero^2+\mu^2},
\eea
where 
\ben
\ba{rcl}
\Di q_0 = \frac{2\pi\Omega}{T},&~~~&\Di\hqzero = 2 \sin \frac{q_0}{2},\\
\\
\Di q_i = \frac{2\pi Q_i}{L},&~~~&\Di\hq_i = 2 \sin \frac{q_i}{2},
\ea
\een
with $0 \le \Omega < T$ and $0 \le Q_i < L$. Then the contribution from
the rainbow diagram is
\begin{multline}
\tG_2(\bp,t) = \tK_0(\bp)^t g^2\left[\frac{1}{L^3T}\sum_{Q_i,\Omega}\sum_{\tau=1}^t
\left(t-\tau\right) \right. \\
\left. \times
\tGamma(\bq,q_0)\left[e^{iq_0}
\frac{\tK_0(\bp-\bq)}{\tK_0(\bp)}\right]^\tau\right]\;.
\label{geom_sum}
\end{multline}

We now let 
\ben
R(\bp,q) = \left[e^{iq_0}\frac{\tK_0(\bp-\bq)}{\tK_0(\bp)}\right],
\een
where $q \equiv (q_0,\bq)$. Then, using Eq. \eqref{G0}, the one-loop rainbow 
diagram correction to the Green function is
\bea
\tG(\bp,t)&=&[1 - g^2 A(\bp,t) t + g^2 B(\bp,t)]G_0(\bp,t) \nn\\ 
\nn\\
&\sim&[1+g^2B(\bp,t)] e^{-g^2 A(\bfp,t) t}G_0(\bp,t),
\eea
and we deduce that $E_0^{(1)}(L,T,t) = g^2A(0,t)$ and $Z_\psi^{(1)}(L,T,t) =
g^2B(0,t)$. Note that both $E_0^{(1)}$ and $Z_\psi^{(1)}$ depend on $t$ but that for 
$t$ sufficiently large both quantities will approach their asymptotic value. 
We then have
\bea
E_0^{(1)}(L,T,t)=-g^2\frac{1}{L^3T}\sum_{Q_i,\Omega}\sum_{\tau=1}^t
\tGamma(\bq,q_0) R(0,q)^\tau  
\label{eq:finiteE0} \\
Z_\psi^{(1)}(L,T,t)=g^2\frac{1}{L^3T}\sum_{Q_i,\Omega}\sum_{\tau=1}^t\tau 
   \tGamma(\bq,q_0) R(0,q)^\tau . \label{eq:finiteZpsi}
\eea
We first consider $E_0^{(1)}(L,T,t)$. We carry out the geometrical sum and find 
\ben
A(\bp,t) = -g^2\frac{1}{L^3T}\sum_{Q_i,\Omega}
\tGamma(\bq,q_0) \frac{R(\bp,q)}{1-R(\bp,q)}(1-R(\bp,q)^t).
\een
For $|R(0,q)| < 1$, for $\bq$, the limits $T \to \infty,\; t \to
\infty$ can be taken. We have that
\bea
\frac{R(\bp,q)}{1-R(\bp,q)} &=& \frac{\tK_0(\bp-\bq)}{e^{i(p_0-q_0)}
- \tK_0(\bp-\bq)}\nn\\
\nn\\
&=&e^{i(p_0-q_0)}\tK_0(\bp-\bq)\tG_0(\bp-\bq),
\eea
where we have used the on-shell condition for the external quark: $\Di e^{ip_0} = \tK_0(\bp)$. In
this limit we find
\begin{multline}
A(\bp,\infty) =  
-g^2\frac{1}{2i\pi L^3}\sum_{Q_i}\int_{|z|=1}\frac{dz}{z}
e^{-i(p_0-q_0)} \\
\times
\tK_0(\bp-\bq)\tG_0(\omega,\bp-\bq)\;,
\label{A_eqn}
\end{multline}
where $\omega = e^{i(p_0-q_0)}$, with $z = e^{-iq_0}$, and the integral is
over the unit circle in the complex $z$-plane. This is the expression for the rainbow diagram 
derived from the NRQCD Feynman rules applicable in the limits $T \to \infty,\; t \to \infty$. 

We conclude that to account for the effect of finite temporal extent of the lattice in the perturbation
theory we make the replacement
\ben
\tG_0(\omega,\bp-\bq) \longrightarrow \tG_0(\omega,\bp-\bq)[1-R(\bp,q)^t]
\een
for the internal quark propagator, and carry out the sums over the discrete
values of $\bq$ and $q_0$. There remains the choice for the value of $t$ in this expression. 
We found that the results were insensitive to this choice as long as $t$ was not close to either 
$0$ or $T$ and so we chose $t=T/2$ for our calculations. $R(\bp,q)$ is computed automatically 
by a numerical search for the poles of the external and internal propagators which gives 
$\tK_0(\bp)$ and $\tK_0(\bp-\bq)$. For $E_0^{(1)}(L,T,t)$ we set $\bp=0$.

The wavefunction renormalisation, $Z_\psi^{(1)}$, is given by 
\begin{multline}
Z_\psi^{(1)}(L,T,t) = ig^2\frac{\partial}{\partial p_0} \frac{1}{L^3T}\sum_{Q_i,\Omega}\sum_{\tau=1}^t 
\tGamma(\bq,q_0) \\ \times
\left[e^{i(p_0-q_0)}\tK_0(\bp-\bq)\right]^\tau\;,
\end{multline}
evaluated on-shell: $\Di e^{-ip_0} = \tK_0(\bp)$. This is the usual formula applied to our
augmented Feynman rule and the derivative is computed using our automated \verb+taylor+ derivative 
procedure.

In some cases we can have $|R| > 1$. This is the situation for some values of $\bq$ given $\bp$ and
certainly occurs in moving NRQCD (mNRQCD) \cite{horgan09}. Because NRQCD is in the Hamiltonian 
formalism the value of $t$ in Eq. (\ref{geom_sum}) is finite and the singularity in the quark
propagator is removable.  The poles in the gluon propagator are at $z = z_{\pm}$ with $|z_\pm|
\gtrless 1$ and $z_+z_- = 1$. Schematically, Eq. (\ref{A_eqn}) takes the form
\ben
A(\bp) = C\sum_{Q_i}\int_{|z|=1}dz\frac{1}{z-z_-}\frac{1}{z-z_+}\frac{z(1-(a/z)^t)}{z-a}
\een 
where $C$ is a constant and $a = \tK_0(\bp-\bq)/\tK_0(\bp)$. The integration contour is $|z| = 1$ and
is determined by the formalism; no distortion is available in the NRQCD evolution to avoid pole crossing. 
However, the singularity at $z=a$ is removable and so there is no
issue of it crossing the contour. The integration is done by Cauchy's theorem at the $z=z_+$ pole, and the 
factor from the geometric summation is then evaluated to be $(1-(a/z_+)^t)$; the need to consider the pole of
order $(t-1)$ at the origin is then avoided. Since $|a| < |z_+|$ the limit 
$t \to \infty$ can now be taken. This corresponds to the usual rule for analytic continuation in the 
calculation of the Feynman diagram where the radius $|z|$ of the contour is increased to avoid crossing by 
the quark pole at $z=a$.

%%%%%%%%%%%%%%%%%%%%%%%%%%%%%%%%%%%%%%%%%%%%%%%%%%%%%%%%%%%%%%%%%
%     BEGINNING OF SECTION
%%%%%%%%%%%%%%%%%%%%%%%%%%%%%%%%%%%%%%%%%%%%%%%%%%%%%%%%%%%%%%%%%

\section{\label{sec:generating_configurations}Generating Configurations and Gauge Fixing}

%%%%%%%%%%%%%%%%%%%%%%%%%%%%%%%%%%%%%%%%%%%%%%%%%%%%%%%%%%%%%%%%%

\subsection{\label{sec:generating_configurations:langevin_markov_chain}Langevin Markov 
Chain Configurations}
Configurations for the Monte Carlo simulations are generated with a Markov
chain that is updated via a Langevin algorithm.
The Langevin method treats the Markov chain as a classical path in phase space, using 
the action as a potential to enforce the Boltzmann distribution.
Using the notation of section \ref{sec:hbeta-method}, the Langevin equation is given by
\begin{equation}
	\frac{ \partial \tU }{ \partial \tau } = -\frac{ \partial S }{ \partial 
	\tU } + \eta,
	\label{eqn:continuum_langevin}
\end{equation}
where $ S $ is the action, $ \eta $ is a random noise term, and $ \tau $ is the distance 
along the path.
Using the Fokker-Plank equation it can be shown that this path will sample 
the configuration space with probability density
\begin{equation}
        P( \tU ) =  e^{  - S[ \tU ] }\;,
        \label{eqn:boltzmann_distribution}
\end{equation}
the Boltzmann distribution, as desired.

As Eq. \eqref{eqn:continuum_langevin} is an initial value problem,
its solution can be approximated via an iterative method, where the derivative on the left hand side is written as a finite difference, 
with step size $ \epsilon $. This introduces step-size errors in the action
so that the distribution that is simulated is altered to
\begin{equation}
        \overline{P}( \tU ) = e^{  - \overline{S}[ \tU,\epsilon ] }\;,
        \label{eqn:effective_boltzmann_distribution}
\end{equation}
where $\overline{S}[\tU, \epsilon]$ is the simulated action which is expansible as
\begin{equation}
        \overline{S}[ \tU,\epsilon ] = S + \epsilon S_1 +  \epsilon^2 S_2 + \cdots\;.
        \label{eqn:langevin_effective_action}
\end{equation}

The step-size errors in Eq. \eqref{eqn:langevin_effective_action}
can be systematically eliminated using higher order approximations to the derivative in Eq. \eqref{eqn:continuum_langevin}. In this work a second order Runge-Kutta algorithm (RK2) 
which eliminates $\mathcal{O}(\epsilon)$ errors was used. This is implemented as a mid-point method 
adapted to diffusion on a group manifold \cite{catterall:1991}.  

Simulations were run with a step size $ \epsilon = 0.2 $. 
Analysis shows that $ \epsilon $ scaling errors are of the order $ \approx 0.05\% $.
Auto-correlation times were measured to be of the order of $ 5 - 10 $ ($ 25 - 50 $ updates), 
for the plaquette and $ 10 - 20 $ ($ 50 - 100 $ updates) for the twisted Polyakov loop \cite{hart04}.
Here $ 100 $ configurations were skipped between measurements.
For each value of $ \beta_{ pl } $ on each lattice size, 32 independent Markov chains 
were generated.
Each chain produced 128 configurations (4096 configuration in total).

%%%%%%%%%%%%%%%%%%%%%%%%%%%%%%%%%%%%%%%%%%%%%%%%%%%%%%%%%%%%%%%%%

\subsection{\label{sec:generating_configurations:gauge_fixing}Gauge Fixing with Twisted Boundaries}

Configurations generated from the Markov chain have the gauge freedom described in Eq. (\ref{eq:gt}).
%\begin{equation}
%	A_{ \mu } ( x ) \to A^{ G }_{ \mu } ( x ) = G ( x ) A_{ \mu } ( x ) G^{ \dagger } ( 
% x ) + i \left( \partial_{ \mu }  G ( x ) \right) G^{ \dagger } ( x )
%	\label{eqn:continuum_gauge_transformation}
%\end{equation}
%on the lattice this corresponds to
%\begin{equation}
%	U_{ \mu } ( x ) \to U^{ G }_{ \mu } ( x ) = G ( x ) U_{ \mu } ( x ) G^{ \dagger } ( 
%x + \hat{ \mu } ).
%	\label{eqn:lattice_gauge_transformation}
%\end{equation}
This can be fixed by the application of a gauge condition.
In this work we wish to fix the configurations to Coulomb gauge.
In the continuum Coulomb gauge is achieved by the the gauge 
transformation that satisfies
\begin{equation}
         \partial_i A^{ g }_i = 0.
	\label{eqn:continuum_Coulomb_gauge_condition}
\end{equation}
On the lattice this corresponds to maximizing the quantity
\begin{multline}
	W[ g ]  =   \sum_{\bsx, i=1 }^{ 3 } \left[g ( \bx ) \, \tU_{ i } ( \bx ) \, 
g^{ \dagger } ( \bx + \be_i )\right.  \\ 
\left.- \frac{ 1 }{ 16 } g ( \bx ) \, \tU_{ i } ( \bx ) \, \tU_{ i } ( \bx  + \be_i ) \, 
g^{ \dagger } ( \bx + 2 \, \be_i )\right], \label{eqn:lattice_Coulomb_gauge_condition}
\end{multline}
with respect to the gauge transform field $g(\bx)$ for each time slice.
This is $ \mathcal{ O } ( a^2 ) $ improved \cite{Lepage:1997id}.
The maximisation is preformed via a conjugate-gradient method, 
using a backtrack line search.
Each time slice is gauge fixed separately.
Errors due to numerical maximisation are estimated to be 
insignificant. %compared to those due to $ \epsilon $.

Fixing to Coulomb gauge leaves an ambiguity, since it is possible to construct an 
additional purely temporal gauge transformation
\begin{equation}
	 \tU_{ 4 } ( \boldsymbol{ x }, t ) \to \tU^{ g^{ ( T ) } }_{ 4 } ( x ) = g^{ ( T ) } ( t ) 
	 \, \tU_{ 4 } ( \bx ) \, g^{ ( T ) \dagger } ( t + 1 ).
\end{equation}
This gauge transformation must obey the twisted boundary conditions
\begin{equation}
	g^{ ( T ) } ( t )  = \Omega_{ i } \, g^{ ( T ) }( t ) \, \Omega^{ \dagger }_{ i },
\end{equation}
for $ i = 1, 2, 3 $. 
The only solutions are
\begin{equation}
	g^{ ( T ) } = \ONE \, z^n,
	\label{eqn:temporal_gauge_transformations}
\end{equation}
for $ n = 0, \dots , N_c $, where $ z $ is given in \eqref{eqn:omegas_and_z}.

After fixing to Coulomb gauge, each time may be in a different gauge.
In order to measure time dependent operators, the time slices must all be in the same gauge.
The gauges are all fixed to be the same as that on the first time slice.
Since the gauge transformation in \eqref{eqn:temporal_gauge_transformations} form a 
group, this is achieved by applying an additional transformation.
The gauge transformation on the first time slice is set to the the unit matrix
\begin{equation}
	g^{ ( T ) }(  t = 0 ) = \ONE.
\end{equation}
The transformations on subsequent time slices are chosen sequentially to maximise
\begin{equation}
	\mbox{Re} \left[ \, \mbox{Tr} \, g^{ ( T ) }( t - 1 ) \, \tU_4 (  \boldsymbol{ 0 },  t - 1 ) \, 
	g^{ ( T ) \dagger }( t )  \, \right],
\end{equation}
for $ t  = 1, \dots T  - 1 $.

%%%%%%%%%%%%%%%%%%%%%%%%%%%%%%%%%%%%%%%%%%%%%%%%%%%%%%%%%%%%%%%%%
%     END OF SECTION
%%%%%%%%%%%%%%%%%%%%%%%%%%%%%%%%%%%%%%%%%%%%%%%%%%%%%%%%%%%%%%%%%

\section{\texorpdfstring{$\MS$}{MS bar} matching formulas}
\label{sec:matching_formulas}
The relation between $\alphav$ and $\alphams$ is given by \cite{Schroder:1998vy,peter97a,peter97b}:
\ben
\alphav = \alphams  (1.0 + c_0 \alphams +c_1 \alphams^2  ).
\een
The coefficients are:
\begin{eqnarray}
 c_0 &=& (a_1 + \beta_0 \log(x))/4\pi \nn \\
 c_1 &=& (a_2 + (\beta_0 \log(x))^2   + ( \beta_1 + 2 \beta_0 a_1  )\log(x)    ) /(4\pi)^2 \nn
\end{eqnarray}
with $\log(x)=0$ since both coupling are evaluated at the same scale
\begin{eqnarray}
\beta_0 &=& 11 -2n_f/3., \\
\beta_1 &=& 2(51 - 19n_f/3),\\ 
a_1 &=& ( 31C_a - 20T_fn_f )/9, \\
a_2 &=&  \left(  \frac{4343}{162}+ 4\pi^2-\pi^4/4 +22 \zeta(3)/3   \right)C_a^2  \\
      &&    - \left(  \frac{1798}{81}+56\zeta(3)/3  \right)C_aT_fn_f  \nn \\ \nn
      &&      - \left(\frac{55}{3}-16\zeta(3)  \right) C_f T_f n_f + \frac{400}{81} T_f^2 n_f^2.
\end{eqnarray}
Note the discrepancy between \cite{Schroder:1998vy} and \cite{peter97a}.

The pole to $\MS$ renormalization is calculated to three loops in 
\cite{Melnikov:2000qh}
\ben
\mpole = Z_M(\mb) \mbmb,
\een 
with
\begin{multline}
Z_M(\mb) = 
1 + \frac{4}{3} \frac{\alphams(\mb) }{\pi}  \\ 
+ \left( \frac{\alphams(\mb)}{\pi}  \right)^2 (-1.0414 n_f + 13.4434)  \\
+ \left( \frac{\alphams(\mb)}{\pi}  \right)^3 (0.6527n_f^2 -26.655n_f + 190.595).
\end{multline}
We actually need the inverse of this series which we define as the 3-loop approximation to $1/Z_M$. With $n_f=3$ this is
\begin{multline}
Z_M^{-1}(\mb) = 
1  - 0.42441318 \alphams  \\
- 0.86542701 \alphams^2
-2.94639 \alphams^3.
\end{multline}

The $\MS$ coupling is discontinuous at quark mass thresholds since the heavy mass quarks are explicitly decoupled by matching to a theory with a different number of flavours.
The formula for matching the $n_f$ theory to a theory with $n_l=n_f-1$ flavours at the threshold is \cite{Chetyrkin:1997sg}
\begin{eqnarray}
\label{eq:alphanl}
\alphams^{(n_l)} = \alphams^{(n_f)} 
\left(
1 + \frac{c_2}{\pi^2}  \left(\alphams^{(n_f)} \right)^2
  + \frac{c_3}{\pi^3}  \left(\alphams^{(n_f)} \right)^3
\right),
\end{eqnarray}
with everything evaluated at the threshold scale of the $n_f$ theory and the coefficients
\begin{eqnarray}
c_2 &=& \frac{11}{72}, \\
c_3 &=& \frac{82043}{27648} \zeta(3) + \frac{564731}{124416} - \frac{2633}{31104}n_l. 
\end{eqnarray}

Crossing thresholds for a running mass in a mass independent scheme gives the same difficulties as the coupling. 
%Chetyrkin et al hep-ph/9708255 (eq 8 and 20) 
The relation between the $n_l$ flavour effective theory and the $n_f$ flavour theory for the $\MSbar$ running mass at the threshold is \cite{Chetyrkin:1997un}
\begin{multline}
m^{(n_l)} = m^{(n_f)}
\left(
1 + \frac{0.2060}{\pi^2} \left(\alphams^{(n_f)} \right)^2 \right.\\
+ 
\left. \frac{(1.8476 + 0.0247n_l)}{\pi^3} \left(\alphams^{(n_f)} \right)^3
\right).
\label{mdec}
\end{multline}
For the inverse of these operations we include higher order terms so that it reproduces the original value to better accuracy.

%%%%%%%%%%%%%%%%%%%%%%%%%%%%%%%%%%%%%%%%%%%%%%%%%%%%%%%%%%%%%%%%%%%%%%%%%%%%%%%%
%%%%%%%%%%%%%%%%%%%%%%%%%%%%%%%%%%%%%%%%%%%%%%%%%%%%%%%%%%%%%%%%%%%%%%%%%%%%%%%%
% References
%%%%%%%%%%%%%%%%%%%%%%%%%%%%%%%%%%%%%%%%%%%%%%%%%%%%%%%%%%%%%%%%%%%%%%%%%%%%%%%%
%%%%%%%%%%%%%%%%%%%%%%%%%%%%%%%%%%%%%%%%%%%%%%%%%%%%%%%%%%%%%%%%%%%%%%%%%%%%%%%%
%\bibliographystyle{h-physrev4}
%%%%%%%%%%%%%%%%%%%%%%%%%%%%%%%%%%%%%%%%%%%%%%%%%%%%%%%%%%%%%%%%%%%%%%%%%%%
\bibliography{mb_v3}
%%%%%%%%%%%%%%%%%%%%%%%%%%%%%%%%%%%%%%%%%%%%%%%%%%%%%%%%%%%%%%%%%%%%%%%%%%%

\end{document}